\documentclass{emulateapj}
\usepackage{graphicx}
\usepackage{natbib}

\newcommand{\simgt}{\lower.5ex\hbox{$\; \buildrel > \over \sim \;$}}
\newcommand{\simlt}{\lower.5ex\hbox{$\; \buildrel < \over \sim \;$}}

\def\btheta{\mbox{\boldmath $\theta$}}

\slugcomment{26, February, 2010, ApJ, Accepted}

\shorttitle{Substructures in COMA Cluster}
\shortauthors{Okabe et al.}

\begin{document}


\title{Weak Lensing Mass Measurements of Substructures in COMA Cluster
with Subaru/Suprime-Cam\altaffilmark{*}}
\altaffiltext{*}{Based on data collected at Subaru Telescope and obtained from
the SMOKA, which is operated by the Astronomy Data Center, National
Astronomical Observatory of Japan. }


\author{N. Okabe\altaffilmark{1,2}}
\email{okabe@asiaa.sinica.edu.tw}

\author{Y. Okura\altaffilmark{2.3}}


\author{T. Futamase\altaffilmark{2}}


\altaffiltext{1}{Academia Sinica Institute of Astronomy and Astrophysics
(ASIAA), P.O. Box 23-141, Taipei 10617, Taiwan}
\altaffiltext{2}{Astronomical Institute, Tohoku University \\
Aramaki, Aoba-ku, Sendai, 980-8578, Japan}
\altaffiltext{3}{ National Astronomical Observatory of Japan, 2-21-1 Osawa, Mitaka, Tokyo 181-8588}


\begin{abstract}
We obtain the projected mass distributions for two Subaru/Suprime-Cam fields in the southwest region
 ($r\simlt 60\farcm$) of the Coma cluster ($z=0.0236$) by weak
 lensing analysis and detect eight subclump candidates.   
We quantify the contribution of background large-scale structure (LSS) on the
 projected mass distributions using SDSS multi-bands and photometric
 data, under the assumption of mass-to-light ratio for field galaxies.
We find that one of eight subclump candidates, which is not associated with any
 member galaxies, is significantly affected by LSS lensing.
 The mean projected mass for seven subclumps extracted from the main cluster
 potential is $\langle M_{{\rm 2D}}^{\rm
(corr)}\rangle=(5.06\pm1.30)\times10^{12}h^{-1}M_\sun$ after a
 LSS correction.
 A tangential distortion profile over an ensemble of subclumps is well described by
 a truncated singular-isothermal sphere model and a truncated NFW model.
 A typical truncated radius of subclumps, $r_t\simeq 35~h^{-1}{\rm
 kpc}$, is derived without assuming any relations between mass and light for member galaxies. 
The radius coincides well with the tidal radius,
 $\sim42~h^{-1}{\rm kpc}$, of the gravitational force of the main cluster. 
Taking into account the incompleteness of data area, a projection
 effect and spurious lensing peaks,
it is expected that mass of cluster substructures account for $19$
 percent of the virial mass, with $13$ percent 
 statistical error.
 The mass fraction of cluster substructures is in rough agreement with numerical simulations. 
\end{abstract}





\section{Introduction}

The cold dark matter (CDM) paradigm predicts the presence of 
numerous substructures in dark halos on any scale,
because less massive objects form earlier and become more massive
through mergers. 
Indeed, high-resolution N-body simulations have shown an assemble history that 
subhalos continually fall into larger halos. 
When interior subhalos penetrate into a central region of a massive parent halo,
subhalos are disrupted by its strong tidal field. 
As a result, the original subhalo mass is reduced by the tidal effect and becomes
a part of smoothed component of the parent halo (e.g De Lucia et al. 2004; Gao et al. 2004). 
Therefore, a study of subhalo properties is of vital importance to understand
an assemble history in halo environments.
Furthermore, a statistical study for subhalos, such as mass function,
would offer a powerful test of the CDM model on scales less than several Mpc.

Gravitational lensing analysis on background galaxies is the unique technique to 
map out mass distributions of any object, such as galaxies and clusters, 
regardless of the dynamical state. 
It therefore enables us to explore substructures in primary halos and to 
measure directly their masses. 
Indeed, galaxy-galaxy lensing studies in clusters (e.g. Natarajan \& Springel 2004; 
Natarajan, De Lucia \&  Springel 2007) revealed cluster substructures
and measured their mass functions 
under the assumption of a scaling relation between mass and light. 
However, a technique requiring no assumption of mass-to-light relation
is of paramount importance, because subhalo size in a strong tidal field
depends on its orbit parameters as well. 
Furthermore, we cannot rule out a possibility that gaseous galaxies in a
gaseous environment of galaxy clusters are offset from subhalo centers
because of ram pressure. 
Even a slight offset prevents us from measuring subhalo masses
 accurately, because a mis-centering of tangential distortion profile causes large error 
 in mass estimations especially within inner regions (e.g Yang et al 2006; Johnston et al. 2007). 
It is therefore of importance to explore subhalos and measure their masses 
based on a weak mass reconstruction technique independent of any mass-to-light scaling relations. 

As demonstrated by Okabe \& Umetsu (2008), a systematic weak lensing study on
seven mering clusters in the range of $z\sim0.055-0.28$ is capable of discovering 
massive substructures associated with cluster major majors. 
However, the limit of angular resolution of reconstructed mass distributions, within the redshift range, 
makes it difficult to discover less massive substructures associated with cluster galaxies. 
On the other hand, it would be easier to detect less massive substructures in lower redshift 
clusters in spite of the weakness of their lensing signal,  
because the number of available source galaxies increases 
thanks to larger apparent size of objects at lower redshift.
Therefore, weak lensing study of low-redshift clusters 
will provides us with a good opportunity to detect and measure smaller subhalo masses in clusters.
As the first step, we select Coma cluster for the target to measure
 subhalo masses by weak lensing analysis alone. 
The redshift of Coma cluster is 0.0236 and is known as one of the most massive clusters near us.
We analyze archival Subaru/Suprime-Cam data (Miyazaki et al. 2002)
to measure subhalo masses 
found in projected mass distributions as well as cluster virial mass, and
calculate the mass fraction of substructures.  
We also investigate lensing from background large-scale structure (LSS)
 in a quantitative way, using SDSS multi-bands and photometric data.

The outline of this paper is as follows. We briefly describe the data
analysis in \S\ref{sec:data} and measure the three-dimensional mass
enclosed within a spherical region of a given radius using a
tangential shear profile in \S\ref{sec:mass}. 
\S \ref{sec:map} represents projected distributions of mass and member
galaxies, quantifies false lensing peaks and estimate 
background lensing effects on the weak lensing mass reconstruction.
In \S \ref{sec:submass}, we measure the two-dimensional masses for
subclumps with and without LSS lensing correction. 
In \S \ref{sec:stack}, we fit a tangential shear profile over an
ensemble of subclump candidates 
and obtain the typical truncated radius and mass of subhalos.
\S\ref{sec:dis} is devoted to the discussion. 
Throughout the paper we adopt cosmology parameters $\Omega_{m0}=0.27$
and $\Omega_{\Lambda}=0.73$. At the redshift of Coma cluster 
$1\farcm=20.0h^{-1}{\rm kpc}$.

\section{Data Analysis} \label{sec:data}

We retrieved two $R_{\rm c}$ image data (Yoshida et al. 2008) from the Subaru archival data
(SMOKA\footnote{http://smoka.nao.ac.jp/index.jsp}). 
Pointings of imaging data are the central region of $r\simlt 30\farcm$
from cD galaxy NGC4874 and the outskirts region of $r\sim30-60\farcm$. They
cover the southwest part of this cluster.
The data were reduced by the same imaging process of using standard pipeline
reduction software for Suprime-Cam, SDFRED \citep{yag02,ouc04}, as
described in Okabe \& Umetsu (2008).
Astrometry and the photometric calibration were conducted using the Sloan
Digital Sky Survey (SDSS) data catalog.
The exposure times are $42$ and $16$ minutes for the central and
outskirt regions, respectively.  
  
Our weak lensing analysis was done using the IMCAT package provided by
N. Kaiser(Kaiser, Squires \& Broadhurst 1995\footnote{http://www.ifa.hawaii/kaiser/IMCAT}).  
We use the same pipeline as Okabe et al. (2009) with some modifications
followed by Erben et al. (2001) (also see Okabe \& Umetsu 2008). 
In the pipeline, we first measure the image ellipticity, $e_\alpha$, from the
weighted quadrupole moments of the surface brightness of each object
and then correct the PSF anisotropy as $e_\alpha'=e_\alpha-P_{\rm
sm}^{\alpha\beta}(P_{\rm sm}*)^{-1}_{\beta\gamma}e^\gamma*$, where $P_{\alpha\beta}$ is the
smear polarizablity tensor and the asterisk denotes the stellar objects.
We fit the stellar anisotropy kernel $(P_{\rm sm}*)^{-1}_{\alpha\beta}e^\beta*$ with the
second-order bi-polynomials function in several subimages whose sizes
are determined based on the typical coherent scale of the measured PSF
anisotropy pattern. We finally obtain the reduced shear $g_\alpha=\gamma_{\alpha}
/(1-\kappa)=(P_g)_{\alpha\beta}^{-1}e_{\beta}'$ using the pre-seeing shear
polarizablity tensor $P_g$. We adopt the scalar value 
$(P_g)_{\alpha\beta}={\rm Tr}[P_g]\delta_{\alpha\beta}/2$, following the technique described in Erben
(2001).

We ran the pipeline for each imaging data and obtained the shear catalogue of
source galaxies whose magnitude ranges are $20-25$ ABmag and
half-light-radius are $\bar{r}_{h}^*+\sigma (r_h^*) < r_h < 10$ pixel,
where $\bar{r}_h^*$ and $\sigma (r_h^*)$ are the mean and $1\sigma$ error
for stellar objects, respectively.
Here the upper limit of magnitude is determined by the outskirt data of short exposure time, 
although faint galaxies in the range of $25-26$AB mag are usable in the data of central region.  
Since apparent sizes of unlensed galaxies, mainly cluster members,
are large in general,
our source galaxy selection efficiently excludes member galaxies which dilutes lensing strengths. 
The number density of source galaxies is $\simeq23{\rm~arcmin}^{-2}$.

\section{Cluster Mass Measurement} \label{sec:mass}

We measure a tangential shear component,
$g_+=-g_{1}\cos2\varphi-g_{2}\sin2\varphi$ and the $45$ degree rotated
component, $g_\times=-g_{1}\sin2\varphi+g_{2}\cos2\varphi$, with respect
to the cluster center, where $\varphi$ is the position angle in counter
clockwise direction from the first coordinate. 
Then, the profiles of shear components $g_\beta=(g_+, g_\times)$ are obtained
from the weighted azimuthal average of the distortion components of source galaxies as
$\langle{g_{\beta}}\rangle(\theta_n)=\sum_{i}u_{g,i}
g_{\beta,i}/\sum_i u_{g,i}$ with a statistical weight
$u_{g,i}=1/(\sigma_{g,i}^2+\alpha^2)$, where 
subscripts 'n' and 'i' denote the n-th radial bin and i-th source object.
We adopt the softening constant $\alpha=0.4$ which is a typical value
of the mean rms ${\bar \sigma}_g$ over source galaxies.  

There are two cD galaxies (NGC 4874 and NGC 4889) in the central region
of Coma cluster. We adopt the center of Coma cluster as NGC 4874 because 
a number of luminous galaxies are concentrated around NGC 4874 in our
optical image, and the peak of X-ray surface brightness is close to
NGC 4874 (see also Figure \ref{fig:xmm}). 
The shear profile covers the range of $4\farcm-60\farcm$ with 5 bins.
It corresponds to the first bin of Kubo et al. (2007). 
We fit the shear profile with the universal profile proposed by Navarro,
Frenk \& White (1996; hereafter NFW profile) and a singular isothermal
sphere (SIS) halo model. 
We assume that the redshift of source background galaxies is $\langle
z_s \rangle=1$. 
An uncertainty of source redshift in mass estimates is
negligible because the lens distance ratio, $D_{ls}/D_s$, at such a low
redshift cluster weakly depends on source redshifts.

The NFW mass model is described by two parameters of 
the three-dimensional mass $M_{\rm NFW}(<r_\Delta)$ 
and the halo concentration $c_{\rm \Delta}=r_{\Delta}/r_s$, where $r_s$ is a scale radius
and $r_\Delta$ is a radius at which the mean density is
$\Delta$ times the critical mass density, $\rho_{\rm cr}(z)$, at the cluster redshift.
The density profile of NFW mass model is expressed in the form of
\begin{eqnarray}
\rho_{\rm NFW}(r)=\frac{\rho_s}{(r/r_s)(1+r/r_s)^2}.
\end{eqnarray}
The three-dimensional cluster mass for NFW mass model is obtained by 
\begin{eqnarray}
M_{\rm NFW}(<r_{\Delta})=
\frac{4\pi\rho_s r_\Delta^3}{c_\Delta^3} m(c_\Delta), \label{eq:Mnfw}
\end{eqnarray}
with
\begin{eqnarray}
m(x)&=&\log(1+x)-\frac{x}{1+x}.
\end{eqnarray}
For reference with other works, results of mass and
concentration parameter within radii for $\Delta=2500,500$ \& $200$ and 
virial overdensity $\Delta=\Delta_{\rm vir}\simeq98$ (Nakamura \& Suto 1997),
are listed in Table \ref{tab:mass}.
The density profile for SIS mass model is given by 
\begin{eqnarray}
\rho_{\rm SIS}(r)&=&\frac{\sigma_{v}^2}{2\pi G r^2}, \label{eq:rhoSIS}
\end{eqnarray}
where the one-dimensional velocity dispersion, $\sigma^2_v$, is a parameter.

The resulting parameters are summarized in Table \ref{tab:mass}.
Since the covering area of the data is small $\simlt 60\farcm$, the NFW mass is not 
constrained well. 
This is why the mass $M_\Delta$ determined by fitting the tangential
shear is sensitive to the tangential distortion at $r_\Delta$ (Okabe et al. 2009).
Since the best-fit virial radius is $\sim98\farcm5$, 
we require data of wider region to measure the cluster mass accurately.
We note that the $\chi^2$ is quite small because the number of
background galaxies is scarce and the intrinsic ellipticity noise is large. 
The covering area of our data is only $\sim16\%$ within $60\farcm$.
If the data covers whole area, the error would improve $\sim2.5$
times and the $\chi^2$ would become close to $1$.
The NFW virial mass changes $-0.5\%$ and $+0.8\%$ if the mean source
redshift is changed to $\langle z_s \rangle =1.2$ and $0.8$, respectively.

We also perform a fitting taking into account a large-scale structure (LSS) lensing
effect. 
The estimation of LSS effect on lensing signal will be described in
detail in \S \ref{subsec:proj}.
The best-fit parameters are summarized in Table \ref{tab:mass2}.
The LSS effect is not significant on cluster mass estimate.

\begin{figure}
\epsscale{1.0}
\plotone{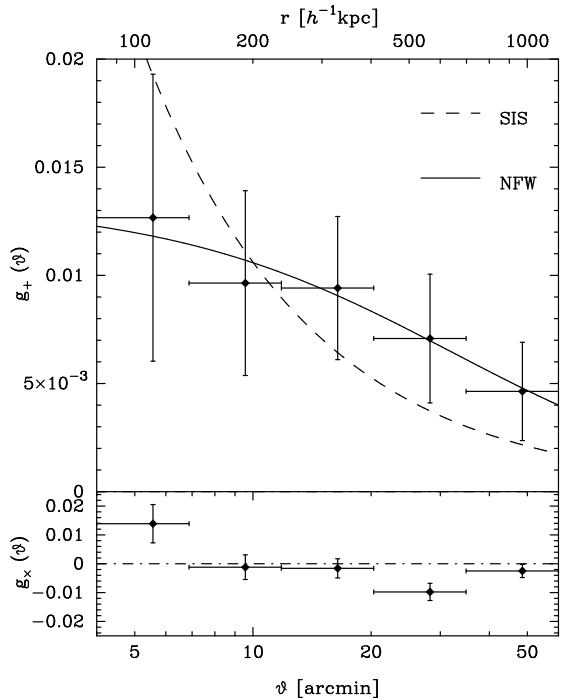}
\caption{Profiles of tangential shear component (top panel), $g_+$, and the $45$ degree rotated
component (bottom panel), $g_\times$. The solid and dashed lines are the best-fit NFW
 and SIS models, respectively}.
\label{fig:g+}
\end{figure}

\begin{table*}
\caption{Best-fit Mass Models without a LSS lensing correction} \label{tab:mass}
\begin{center}
\begin{tabular}{ccccc}
\hline
\hline
NFW & $\chi^2/{\rm d.o.f}=0.1/3$ 
    &
    & 
    &  \\
\hline
$\Delta$ &
$M_{\rm NFW}(<r_\Delta)$ &
$c_{\Delta}$ &
$r_{\Delta}$ &
$\theta_{\Delta}$   \\
 & 
$10^{14} h^{-1} M_\sun$ &
 &
$h^{-1}$kpc &
arcmin         \\
(1) & (2) & (3) & (4) & (5) \\
\hline
vir &
$8.85_{-5.12}^{+19.87}$ &
$3.49_{-1.78}^{+2.55}$  & 
$1972.2_{-493.7}^{+947.7}$  &
$98.30_{-24.54}^{+47.19}$   \\
$200$ & 
$6.56_{-3.60}^{+11.93}$&
$2.49_{-1.34}^{+1.93}$&
$1403.7_{-326.8}^{+579.2}$ &
$70.15_{-16.33}^{+28.94}$  \\
$500$ &
$4.05_{-2.00}^{+5.01}$&
$1.53_{-0.90}^{+1.34}$&
$880.7_{-177.6}^{+270.5}$ &
$44.01_{-8.87}^{+13.52}$  \\
$2500$ & 
$1.18_{-0.42}^{+0.57}$&
$0.61_{-0.40}^{+0.67}$&
$341.2_{-46.3}^{+47.5}$ &
$17.05_{-2.31}^{+2.38}$  \\
\hline
\hline
SIS & $\chi^2/{\rm d.o.f}=4.4/4$ 
    & 
     \\
\hline
$\sigma_v$ (km/s) &
&
& \\
(6) & & \\
\hline
$671.1_{-69.5}^{+73.4}$ & 
&
&
& \\
\hline
\end{tabular}
\tablecomments{
Col. (1): Over-densities $\Delta=\Delta_{\rm vir}, 200, 500$, \& $2500$.
Col. (2): Best-fit three-dimensional cluster mass for NFW mass model within
 $r_\Delta$ at which the mean interior density is
$\Delta$ times the critical mass density, $\rho_{\rm cr}(z)$, at the
 cluster redshift (eq. \ref{eq:Mnfw}).
Col. (3): Best-fit concentration parameter $c_\Delta$.
Col. (4): Radius $r_\Delta$ corresponding to the mass $M_{\rm NFW}(<r_\Delta)$.
Col. (5): Angular size of radius $r_\Delta$
Col. (6): Best-fit velocity dispersion for SIS model (eq. \ref{eq:rhoSIS}). 
The $\chi^2/{\rm d.o.f}$ is the chi-square for best-fits and the degree-of-freedom.
}
\end{center}
\end{table*}

\begin{table*}
\caption{Best-Fit Mass Models with a LSS lensing correction (see
 \S\ref{sec:mass} \& \S\ref{subsec:proj})} \label{tab:mass2}
\begin{center}
\begin{tabular}{ccccc}
\hline
\hline
NFW & $\chi^2/{\rm d.o.f}=0.1/3$ 
    &
     \\
\hline
$\Delta$ &
$M_{\rm NFW}(<r_\Delta)$ &
$c_{\Delta}$ &
$r_{\Delta}$ &
$\theta_{\Delta}$  \\
 & 
$10^{14} h^{-1} M_\sun$ &
 &
$h^{-1}$kpc &
arcmin  \\
(1) & (2) & (3) & (4) & (5) \\
\hline
vir &
$8.92_{-5.17}^{+20.05}$ &
$3.50_{-1.79}^{+2.56}$  & 
$1972.2_{-493.7}^{+947.7}$  &
$98.56_{-24.67}^{+47.36}$                \\
$200$ & 
$6.61_{-3.63}^{+12.06}$&
$2.50_{-1.34}^{+1.94}$&
$1407.7_{-328.4}^{+581.6}$ &
$70.35_{-16.41}^{+29.07}$  \\
$500$ &
$4.09_{-2.01}^{+5.07}$&
$1.57_{-0.90}^{+1.34}$&
$883.4_{-178.0}^{+272.0}$ &
$44.15_{-8.90}^{+13.59}$  \\
$2500$ & 
$1.19_{-0.42}^{+0.58}$&
$0.61_{-0.40}^{+0.67}$&
$342.5_{-45.0}^{+47.7}$ &
$17.11_{-2.25}^{+2.38}$  \\
\hline
\hline
SIS & $\chi^2/{\rm d.o.f}=4.4/4$ 
    & 
     \\
\hline
$\sigma_v$ (km/s) &
&
& \\
(6) & & \\
\hline
$673.0_{-69.7}^{+73.7}$ &
&
&
& \\
\hline
\end{tabular}
\tablecomments{
See the caption of Table \ref{tab:mass}.
}
\end{center} 
\end{table*}

\section{Projected Distributions of Mass and Galaxies}  \label{sec:map}

\subsection{Mass, Number Density and Luminosity Maps } \label{subsec:massmap}

We reconstruct the lensing convergence field, $\kappa$, from the shear field, 
using the Kaiser \& Squires inversion method (Kaiser \& Squires 1993), following Okabe \& Umetsu (2008).
In the map making, we pixelize the shear data into a pixel grid using a Gaussian
 smoothing kernel $w_g(\theta)\propto \exp[-\theta^2/\theta_g^2]$ and a
 statistical weight $u_{g,i}$ (\S \ref{sec:mass}). 
The shear field at a given position ($\btheta$) is obtained by 
$\bar{\gamma}_{\alpha}(\btheta) = \sum_i w_g(\btheta-\btheta_i) u_{g,i} \gamma_{\alpha,i}/ \sum_i
 w_g(\btheta-\btheta_i) u_{g,i}$, where the weak limit  $g_{\alpha}
 \approx \gamma_{\alpha}$ is assumed. 
We employ the smoothing ${\rm FWHM} = \sqrt{4\ln{2}}  \theta_g=2\farcm 00$. 
The error variance for the smoothed shear is given as
$\sigma^2_{\bar{g}}(\btheta) = 
(\sum_i w_g(\btheta-\btheta_i)^2 u_{g,i}^2 \sigma^2_{g,i})/
\left( \sum_i w_g(\btheta-\btheta_i) u_{g,i} \right)^2$.
We have used $\langle g_{\alpha,i}\, g_{\beta,j}\rangle = 
(1/2)\sigma_{g,i}^2\delta^{\rm K}_{\alpha\beta}\delta^{\rm K}_{ij}$
with $\delta^{\rm K}_{\alpha\beta}$ and $\delta_{ij}^{\rm K}$ being the
Kronecker's delta.
In the linear map-making process, the pixelized shear field is weighted
by the inverse of the variance.

The resulting E-mode and B-mode map of lensing fields are
shown in panels B and C in Figures \ref{fig:coremap} and
\ref{fig:submap} which cover the central region ($26\farcm \times 26\farcm$) 
and the outskirts ($21\farcm\times 21\farcm$), respectively.
Contours are spaced in a unit of $1\sigma$ reconstructed errors.
As seen in panels A and B, we find eight candidates of mass clumps
whose significance is over $3\sigma$ level. 
The panel A in Figures \ref{fig:coremap} and \ref{fig:submap} show the Subaru $R_{\rm
c}$-band images overlaid with contours of reconstructed mass distributions.
We labeled subclumps as shown in panel A.
The significance levels of mass clumps are lower than
those of other clusters at redshift range $z\sim0.055-0.28$ (Okabe \&
Umetsu 2008). 
The B-mode map (panel C) in the central region shows 2 clumps over
$3\sigma$ close to clump candidates 4 and 2 ($3.5\sigma$ and $3.9\sigma$).

We retrieve the SDSS DR7 catalogue (Abazajian et al. 2009) from SDSS CasJobs
site\footnote{http://casjobs.sdss.org/} in order to investigate member galaxy
distributions. We select bright member galaxies by criteria of $i'<19$ ABmag
and $|(g'-i') - (-0.05i'+2.04)|<0.14$, where we use {\it psfMag} for magnitude 
and {\it modelcolor} for color.
We convert from apparent to absolute magnitudes by 
using the k-correction for early-type galaxies under the assumption that
all member galaxies are located at a single cluster redshift.
The galaxy luminosity and density projected distributions are obtained
using the same kernel of weak lensing mass reconstruction.
The overall mass distribution appears to be similar to the galaxy
luminosity and density distributions. 
In particular, the 6 out of 8 mass candidates, but for the clumps 3 and 5, host bright galaxies. 
At the clump candidate 3, groups of faint member galaxies were known (Conselice
\&  Gallagher 1999), while any galaxy group are not found at the candidate 5
region.
We list the luminous galaxy associated with each candidate in Table \ref{tab:massclump}.
We do not always detect mass structures for all known groups or
luminous galaxies.
There are three possibilities for this.
First, the large-scale-structure (LSS) lensing effect prevents us from detecting lensing signals.
Second, dark matter halos associated with almost member galaxies
are less massive than the detection limit ($3\sigma$), $3\times(\pi
\theta_g^2 \Sigma_{\rm cr} \delta \kappa)\sim 3\times
10^{12}h^{-1}M_\sun$.
Third, dark halos lost their mass by the tidal force of the
main cluster and then are smoothly distributed within the smoothing scale of
mass reconstruction.

\subsection{Bootstrap Re-sampling Mass Reconstructions} \label{subsec:boot}

We run 1000 bootstrap simulations for making mass reconstruction in order to
investigate the realization of mass clumps.
In each reconstruction, we generate a bootstrap data-set by 
choosing randomly galaxies, with replacement, from the original shear catalogue 
and then identify mass clumps whose significance level is more than
$3\sigma$.
Figure \ref{fig:bootstrap} shows the resulting distributions of histogram of the
appearance of mass peaks 
These distributions are well associated with mass clumps.
The radii at which $68\%$ of the centroid positions contain are $0\farcm8-3\farcm$. 
Therefore, the detected lensing peaks are realized well in the shear catalogue.

\subsection{Monte-Carlo Realizations} \label{subsec:monte}

\begin{figure*}
\plottwo{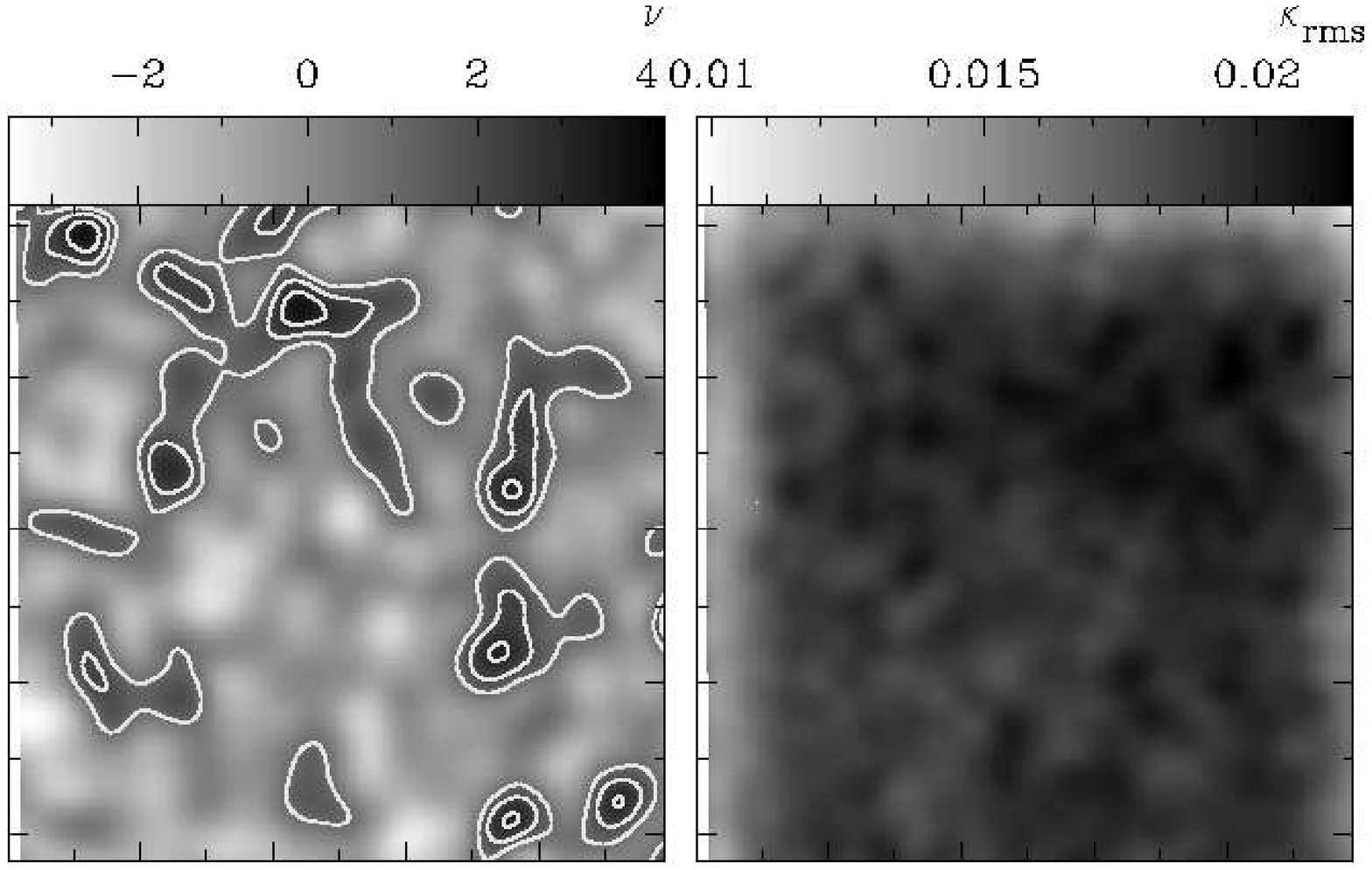}{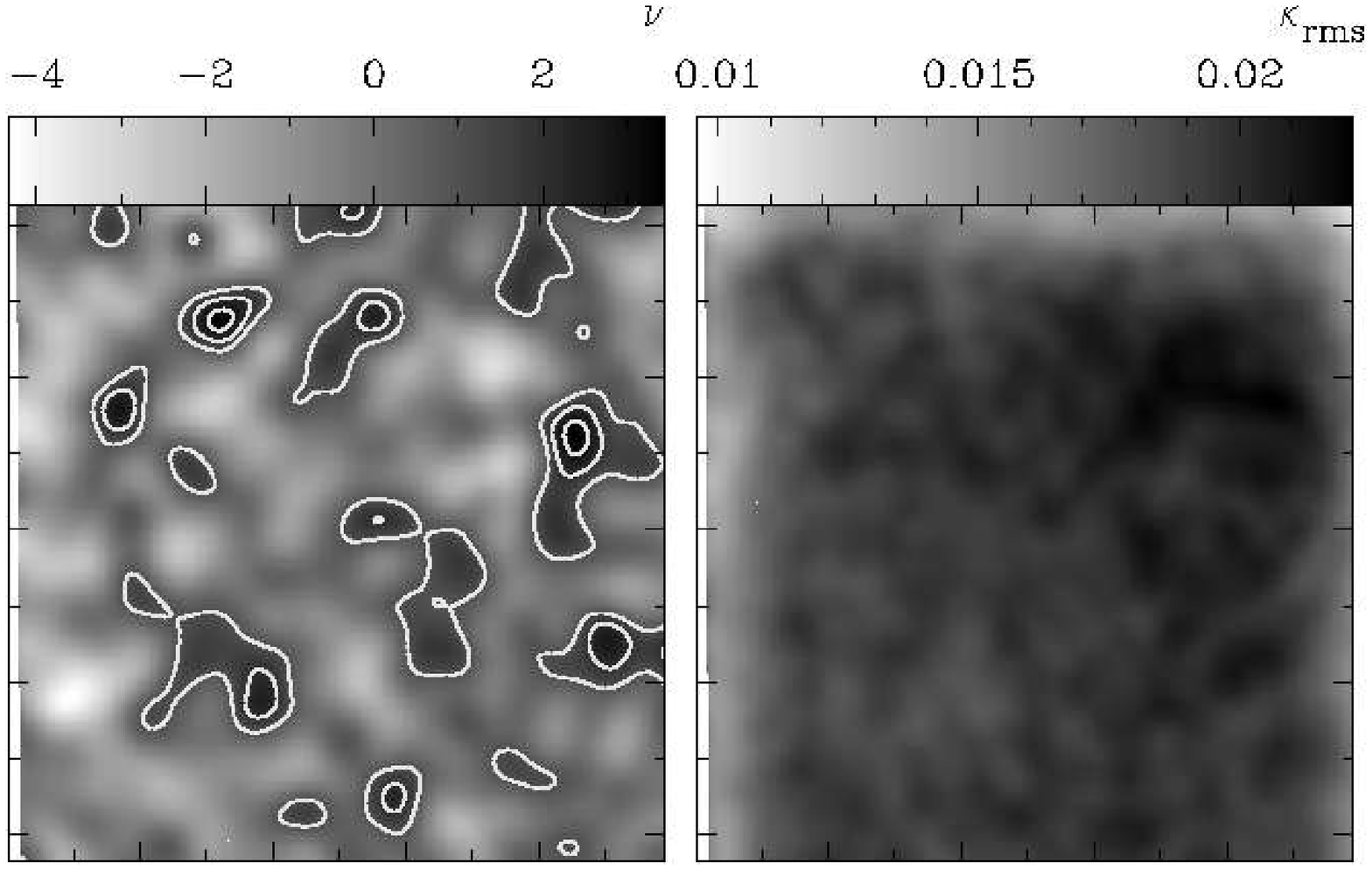}
\hspace{15mm}
\plottwo{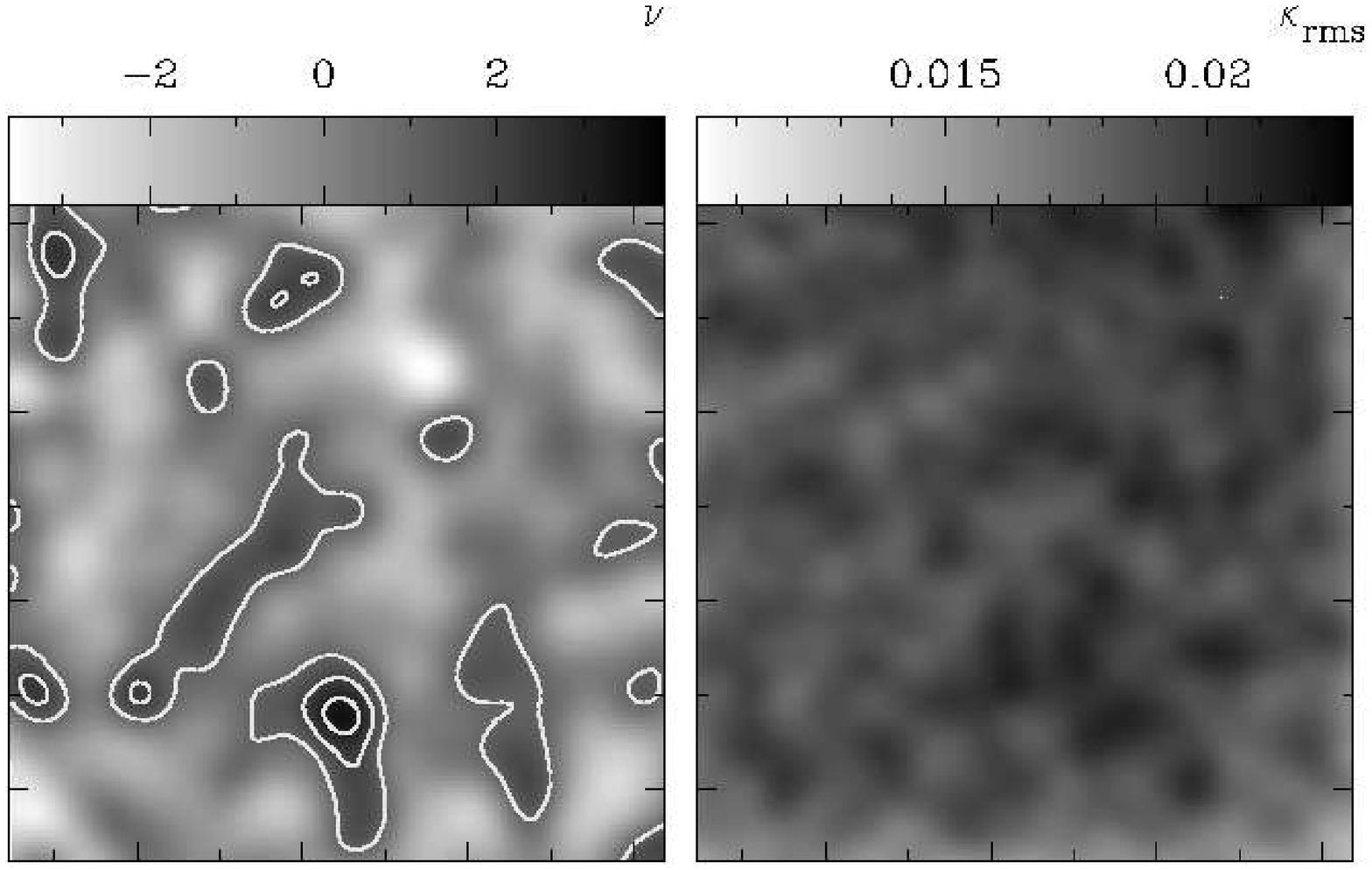}{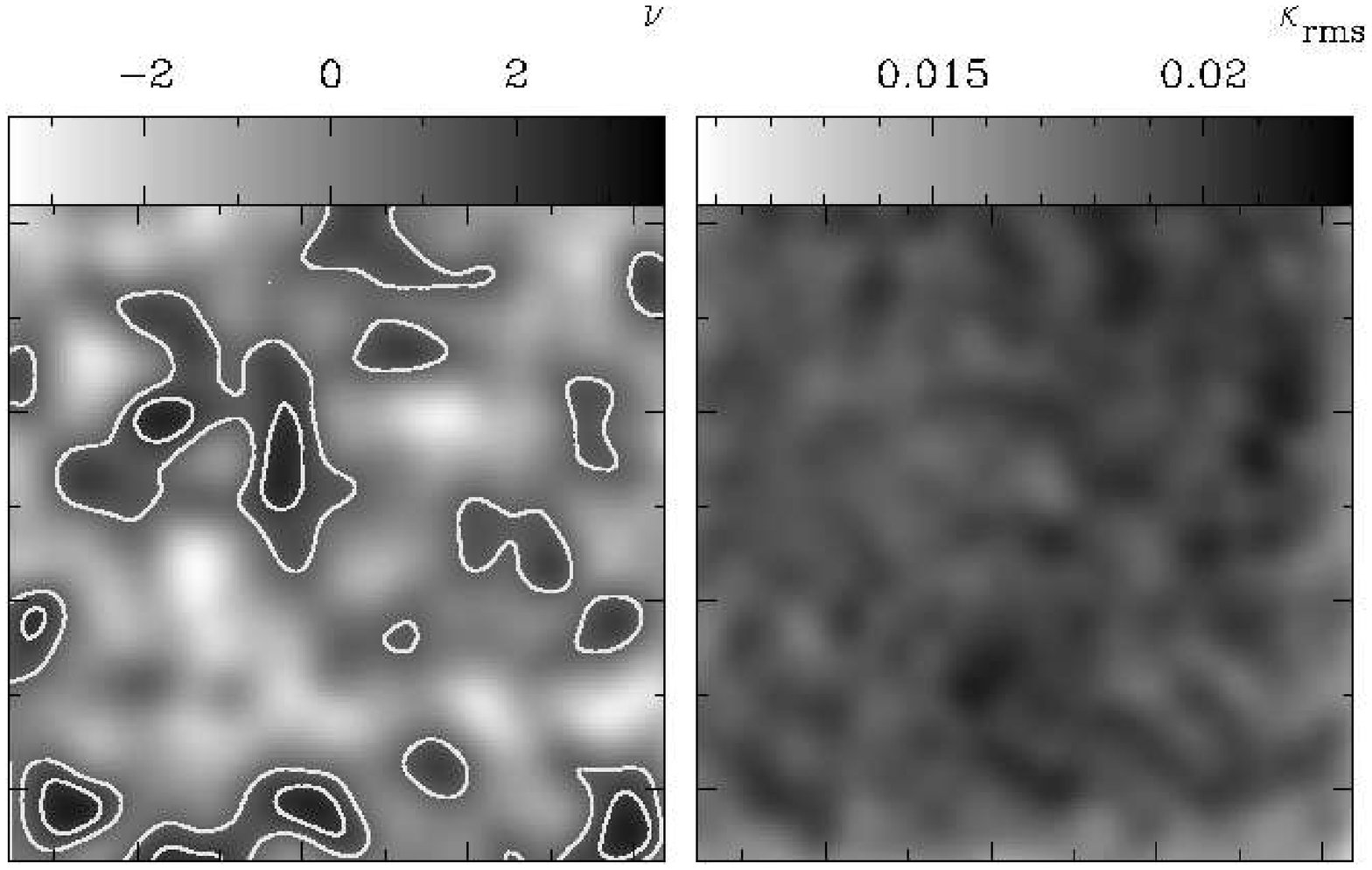}
\caption{{\it Top-left}: The maps for significance
 $\nu=\kappa/\kappa_{\rm rms}$ (left) and noise $\kappa_{\rm rms}$
 (right) for E-mode in the central region, based on 1000 Monte-Carlo
 realizations. The contours are spaced by $\nu=1,2,$ and $3$.
{\it Top-right}: The same maps for B-mode in the central region as the top-left panel.
{\it Bottom-left}: The same maps for E-mode in
 the outskirts.
{\it Bottom-right}: The same maps for B-mode in
 the outskirts.
}
\label{fig:numap}
\end{figure*}

We next construct a noise map, $\kappa_{\rm rms}$, from $1000$
Monte-Carlo realizations, following Miyazaki et al. (2007).
The position and shear components of background galaxy catalogue are
randomly shuffled in each realization.
A mass map for a new background catalogue is reconstructed by applying
the same procedure as making the original $\kappa$ maps.
We estimate the rms noise in each pixel and make the noise maps,
$\kappa_{\rm rms}$ for the central region and the outskirts.
Noise maps are not changed even if we randomly choose half
of catalogue. 
The significance maps, $\nu=\kappa/\kappa_{\rm rms}$, are
obtained by dividing the original $\kappa$ maps by the $\kappa_{\rm
rms}$ map. 
The resulting $\nu$ and $\kappa_{\rm rms}$ maps for both E- and B-modes
 are shown in Figure \ref{fig:numap}.
The variation along the left and top edges of $\kappa_{\rm rms}$ map in
the central region is smaller than the other region. 
This is why fewer galaxies exist around the boundary of the optical image. 
The significance maps, $\nu$, are consistent with the original E-
and B-mode maps.
The significance for subclump candidates, $\nu$, is also consistent with
the original $S/N$ ratio (Table \ref{tab:massclump}).

\begin{figure}
\plotone{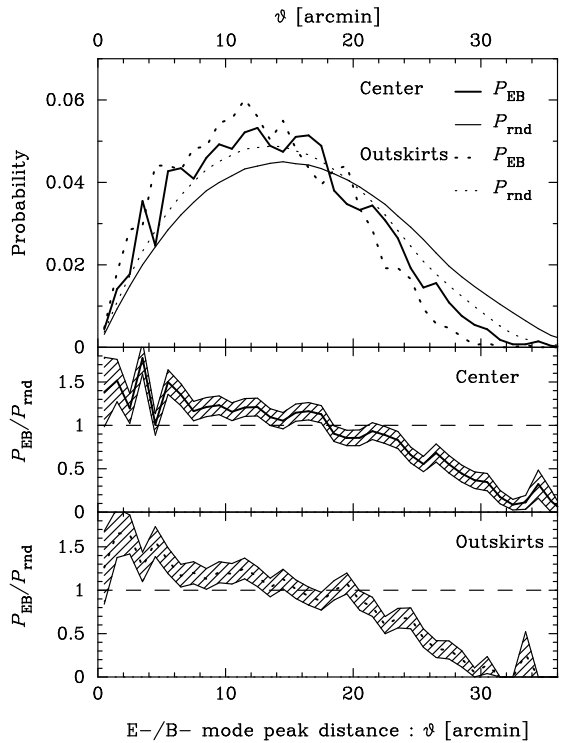}
\caption{{\it Top}: The appearance probabilities as a function of
 distance between E- and B-mode peaks for the central region and the
 outskirts. $P_{\rm EB}$ is the appearance probability obtained by
 1000 Monte-Carlo realizations. $P_{\rm rnd}$ is the probability that E- and
 B-mode peaks are randomly and independently appeared.
{\it Middle}: The ratio of $P_{\rm EB}/P_{\rm rnd}$ for the central
 region.
{\it Bottom}: The ratio of $P_{\rm EB}/P_{\rm rnd}$ for the outskirts.
}
\label{fig:Peb}
\end{figure}

Are positions of E-mode and B-mode peaks correlated ?
In the central region, two B-mode peaks whose significance level is above
$3\sigma$ are appeared close to E-mode peaks.
We calculate the probability, $P_{\rm EB}$, as function of the distance between E-
and B-mode peaks appeared in Monte-Carlo realizations.
The following result does not change even when we use half of
realization data.
Since the appearance probability is proportional to the area size, 
we also compute the probability, $P_{\rm rnd}$,
that E- and B-mode peaks are randomly and independently located in each pixel.
Figure \ref{fig:Peb} shows the appearance probabilities, $P_{\rm EB}$, for the central region and
the outskirts, which roughly agrees with the probability of white noise case.
A slight excess of the ratio $P_{\rm EB}/P_{\rm rnd}$ is found in the
range of $\theta<10\farcm$, but the probability is quite small.
Since they are not significant, we cannot identify fake E-mode peaks using the
distance from B-mode peaks.
The probabilities of large distance $>20\farcm$ is smaller than the
unity, because few peaks are appeared around the edge.
Indeed, the appearance probability in one pixel within $2\farcm$ width
from the boundary is about one-thirds of that in the rest region.
The probability of spurious lensing peaks will be evaluated considering
the large-scale structure lensing effect \S \ref{subsec:spurious}.

\begin{figure*}
\epsscale{1.0}
\plotone{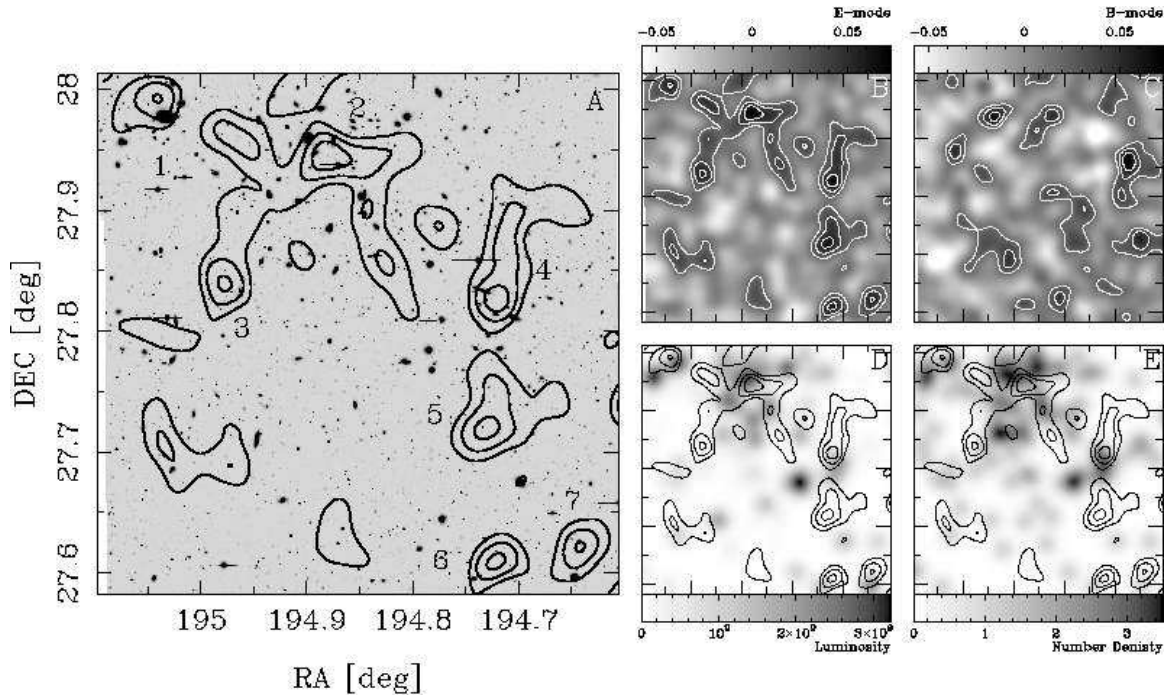}
\caption{The panel A is Subaru $R_{\rm c}$-band image of the central $26\farcm \times
 26\farcm$ cluster region. Two cD galaxies (NGC 4874 and NGC 4889) are
 located around the northeast boundary.
Overlaid are contours of the reconstructed projected mass distribution,
 spaced in a unit of $1\sigma$ reconstruction error ($\delta \kappa=0.018$). The Gaussian FWHM
 is $2\farcm 00$. 
The identified subclumps are labeled in the panel A.
The panels B and C are the lensing $\kappa$ (E-mode) and B-mode fields.
The panels D and E are cluster luminosity and density distributions 
in SDSS $i'$-band smoothed to the same angular resolution of the mass
 map, respectively. Seven clump candidates are found in the central region.
}.
\label{fig:coremap}
\end{figure*}

\begin{figure*}
\epsscale{1.0}
\plotone{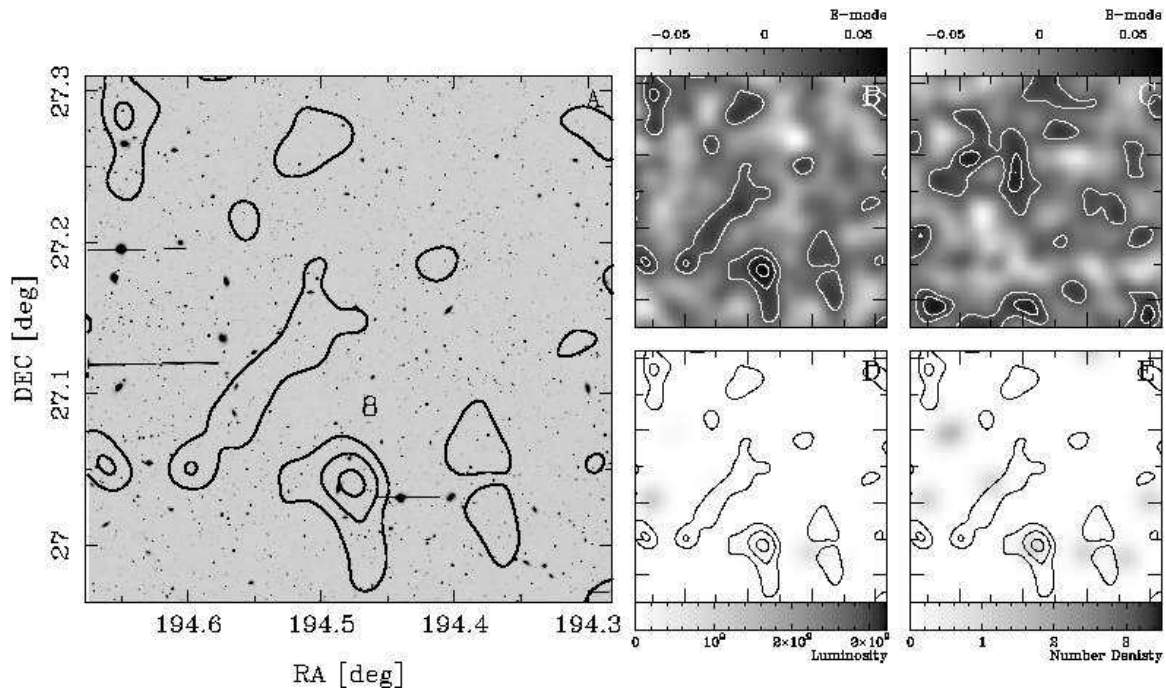}
\caption{The panel A is Subaru $R_{\rm c}$-band image of the outskirts $21\farcm \times
 21\farcm$ cluster region. 
Overlaid are contours of the reconstructed projected mass distribution,
 spaced in a unit of $1\sigma$ reconstruction error ($\delta \kappa=0.020$). 
The Gaussian FWHM is $2\farcm 00$. 
The identified subclumps are labeled in the panel A.
The panels B and C are the lensing $\kappa$ (E-mode) and B-mode fields.
The clump 8 is far $\sim58\farcm$ from cD galaxy NGC4874. 
The panels D and E are cluster luminosity and density distributions 
in SDSS $i'$-band smoothed to the same angular resolution of the mass
 map, respectively.
One clump candidate is found in the outskirts.
}.
\label{fig:submap}
\end{figure*}

\begin{figure*}
\epsscale{1.0}
\plotone{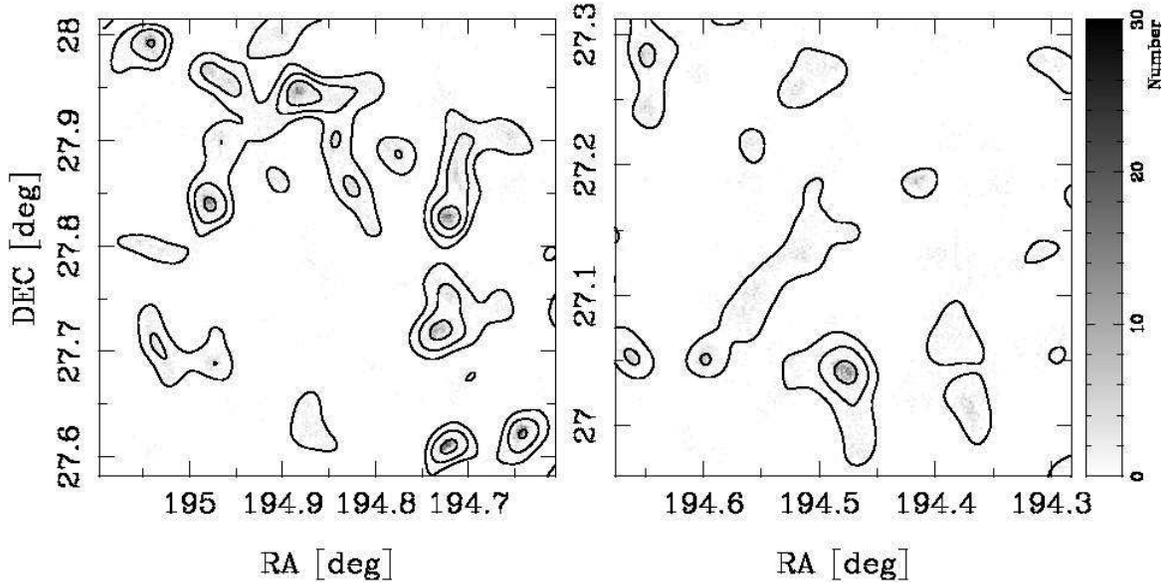}
\caption{The histogram maps of centroid of mass peaks, whose significance level is above $3\sigma$, 
appeared in 1000 bootstrap re-sampling mass-reconstructions.
Contours are the same as Figures \ref{fig:coremap} and \ref{fig:submap}. 
Left and right panels are the central and outskirts regions, respectively. }
\label{fig:bootstrap}
\end{figure*}

\subsection{Projection Effects} \label{subsec:proj}

Since Coma cluster is quite close to us, we cannot rule out a
possibility that lensing signals by background structures significantly contribute to 
the observed ones. 
In this subsection, we quantify the projection effect by background
structures on local convergence peaks appeared in the mass maps,
based on the observational data, rather than a theory.
In this paper, we use the shear catalogue derived from one pass-band data alone,  
which makes quite difficult to measure the contributions of background structures on
lensing signals.

The SDSS DR7 data (Abazajian et al. 2009), on the other hand, allows us
to quantify the contribution, because a huge multi-band data with
photometric redshifts are available. 
We retrieved the data in the region of $10\degr\times 10\degr$ ($190\degr\le {\rm RA}\le 200\degr$ 
and $23\degr\le {\rm DEC}\le 33\degr$). 
Since there is no candidate for galaxy clusters or groups at higher redshift
in the Subaru data field by visual checks, at least, we expect to ignore contributions of background
clusters/groups in our data field. 
We quantify the projection effect by field galaxies with photometric 
catalogue under the assumption of mass-to-light scaling relations (Guzik \& Seljak 2002). 
First, we select galaxy catalogue by $r'<21$, and $z_{\rm
ph}-z_{l}>\delta z= \sigma_{v,max}(1+z_l)/c\simeq0.01$, taking into account the uncertainty of the
photometric redshift due to line-of-sight velocities of member galaxies. 
Here, $z_{\rm ph}$ is the photometric redshift of each galaxy, $c$ is the light
velocity and $\sigma_{v,max}=3000$ is the maximum of the line-of-sight
velocity (Rines et al. 2003). 
The following results do not change even
when we choose the redshift range of $0.5 \delta z$ and $2 \delta z$,
because a relative contribution of low redshift galaxies in the lensing signal is quite small.
The resulting galaxy catalogue has 
a peak at $z_{\rm ph}\sim 0.5-0.6$ in the histogram of photometric redshifts for faint galaxies ($20<r'<21$) in DR6 data (Oyaizu et al. 2008),
If a spectroscopic data of a galaxy is available, we utilize a spectroscopic
redshift instead of photometric one.
Next, we calculate the multi-band luminosities (u'g'r'i'z') within a radius of
$1.93\deg$ from each position of galaxy in faint shear catalogue, corresponding
to $30~{\rm Mpc}$ at $z=0.5$, in order to consider contributions from
unknown clusters/groups around $z=0.5$,
because the two-halo term in the tangential shear measurements (Seljak
2000; Mandelbaum et al. 2005) are dominated around a few tens of Mpc
(Johnston et al. 2007). 
Third, we calculate individual galaxy masses from the multi-band luminosities
(u'g'r'i'z') assuming the mass-to-light ratios obtained by SDSS bands
(Guzik \& Seljak 2002).  
The assumed mass-to-light ratio is in agreement with results in other
bands (Hoekstra et al. 2005). 
Since we adopt the mass and luminosity scaling relation
for a galaxy, a mass of an overdensity region at which 
a distribution of galaxies is concentrated might be overestimated. 
Finally, we assume the mass-concentration relation (Duffy et al. 2008)
 to estimate NFW shear signals at each galaxy position in
 background shear catalogue, and add them all up.
The luminosity scaling relations in multi-bands are complementary with each other to
calibrate the lensing signals from background large-scale structures.
If the assumed mass-luminosity relation is adequate, 
the reduced shears, $g_\alpha$, estimated in each band should coincide with
those in other bands.
The resulting shears in u'g'i' bands are systematically inconsistent with all other
bands, while the shears in r'z' bands have a tight correlation.
We hereafter adopt $g_\alpha^{\rm (LSS)}=(g_\alpha^{\rm (r')}+g_\alpha^{\rm (z')})/2$ as
a model of lensing signals from background large-scale structures.

\begin{figure*}
\epsscale{1.0}
\plotone{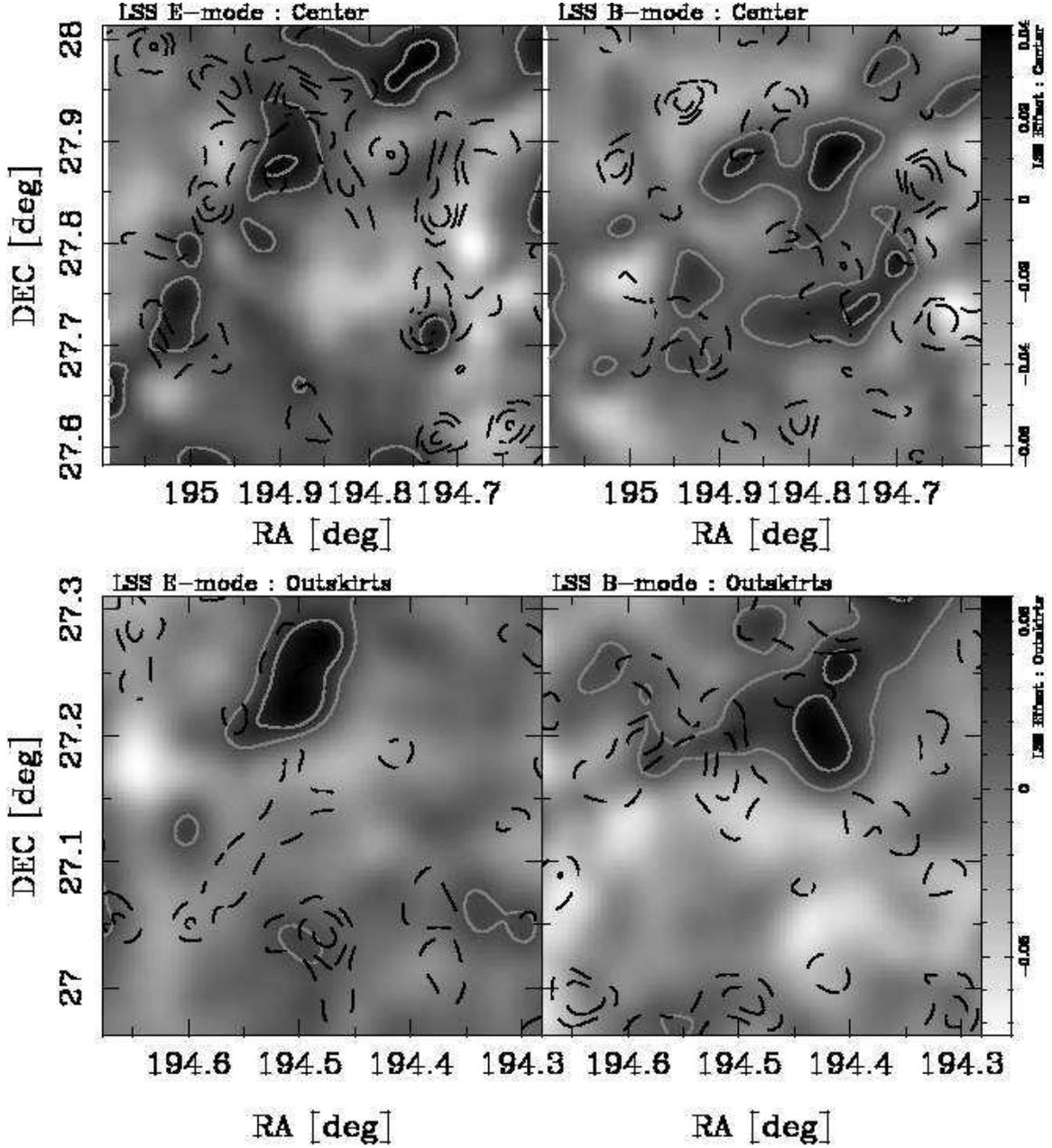}
\caption{The E- and B-mode maps reconstructed by background LSS model
 alone. 
{\it Top-left }: E-mode map in central
 region. {\it Top-right }: B-mode map in central region.  {\it
 Bottom-left }: E-mode map in outskirts. {\it Bottom-right}: B-mode map in outskirts. 
The box sizes and contour scales are the same as those in Figure
 \ref{fig:coremap} and \ref{fig:submap}. 
Gray solid and black dashed lines show contours derived from
 LSS and observed shears, respectively.
There are three LSS structures with a size of a few arcmins at a $\sim2\sigma$ level.
In particular, the LSS convergences at a $\sim1\sigma$ level are found in the clump candidates 5 and 8.
}
\label{fig:LSSmap}
\end{figure*}

We reconstruct the lensing convergence fields with the same kernel of
mass maps (\S \ref{subsec:massmap}) using LSS contributed shears alone. 
Here, we do not add intrinsic shape noises for galaxies as well as
shears from main cluster, in order to investigate the LSS lensing effect only. 
Figure \ref{fig:LSSmap} shows the resulting E- and B-mode maps.
The signal-to-noise ratio for background LSS convergence field is at
$\sim 2 \sigma$ level.
In subclump candidates 5 and 8, peaks of $\sim 1 \sigma$ are found in the
LSS field, while 
no galaxy concentration in SDSS catalogue is found there.
It indicates a possibility that an appearance of two clumps in the mass
maps is biased by a LSS lensing effect.

\subsection{Probability of Spurious Lensing Peaks} \label{subsec:spurious}

We next investigate a probability to detect spurious lensing peaks
by a composition of the LSS and main cluster lensing signals and intrinsic shapes. 
We consider shears composed of $g_\alpha=g_\alpha^{\rm (main)}+g_\alpha^{\rm (LSS)}+e_\alpha^{\rm (int)}$,
where $g_\alpha^{\rm (main)}$ is a best-fit NFW model in \S
\ref{sec:mass} and $e_\alpha^{\rm (int)}$ is an intrinsic shape.
We produce the intrinsic ellipticities with a Gaussian distribution whose
mean value is $|e_\alpha^{\rm (int)}|=0$ and the standard deviation is
$|\delta e_\alpha^{\rm (int)}|=0.28$, corresponding to observed shear distributions.
We then reconstruct mass maps and identify the mass peaks above
$3\sigma$ level, as the same procedures (\S\ref{subsec:massmap}).
We repeat this 1000 times.
The probability, $P_{\rm spur}^{\rm (LSS)}$, to detect spurious lensing
peak within a smoothing scale of each mass clump
candidate except main cluster center is summarized in Table
\ref{tab:massclump}. 
The probabilities for clump candidates 5 and 8 are $P_{\rm spur,5}^{\rm
(LSS)}=12.2\%$ and $P_{\rm spur,8}^{\rm (LSS)}=1.8\%$, respectively. 
They are much higher than those for the other clump candidates $P_{\rm
spur}^{\rm (LSS)}<0.4\%$.
We perform the same steps without LSS lensing signals ($g_\alpha=g_\alpha^{\rm
(main)}+e_\alpha^{\rm (int)}$). 
The probabilities for candidates 5 and 8 are $P_{\rm spur,5}^{\rm (noise)}=0.7\%$
and $P_{\rm spur,8}^{\rm (noise)}=0.9\%$, respectively.
The probability to count spurious peaks in candidate 5 region becomes
significantly higher by LSS lensing effect.
We also compute the averaged probability in the region excluding clump candidates
in the Monte Carlo simulation taking into account the main cluster and intrinsic noises.
The resulting averaged probabilities are $\langle P_{\rm spur}^{\rm (noise)}\rangle=0.30~{\rm
and}~~0.22\%$ in central and outer regions, respectively.
Here, we assume that the realization for spurious peaks is Possionian. 
The probabilities for candidate 5 and 8 regions are still higher even
except LSS effects. 
It might be due to distributions for sheared galaxies.
We define the {\it bias} for a preference to detect  
a clump in weak mass reconstruction 
as $b_{\rm map}=P_{\rm spur}^{\rm (LSS)}/\langle P_{\rm spur}^{\rm
(noise)}\rangle$. As summarized in Table \ref{tab:massclump}, the bias
for candidates 5 and 8 are significantly higher than those for others.
The appearances for candidates 5 and 8 in the reconstructed mass maps are likely to be due to
background LSS lensing distortions, at least partially.
In the next two sections, we will quantify this more accurately based on two complementary mass
measurements using shears, because each pixel in the convergence field is
correlated by smoothing kernel.


\section{Projected Mass Measurements for Subclumps }  \label{sec:submass}

We measure projected mass of subclump candidates (labeled in Figures
\ref{fig:coremap} and \ref{fig:submap}) which are identified above the
$3\sigma$ significant level in the mass map.
The projected masses, $M_{\rm \zeta_c}(<\theta)$, are estimated by 
the so-called $\zeta_c$-statistics (Clowe et al. 2000) which is
modified version of original one (Fahlman et al. 1994),
\begin{equation}
M_{\rm \zeta_c}(<\theta) = \pi{\theta}^2\Sigma_{\rm cr}\zeta_c(\theta,\theta_{{\rm
 b1}},\theta_{{\rm b2}}),
\end{equation}
where,
\begin{eqnarray}
\zeta_c(\theta,\theta_{{\rm b1}}, \theta_{{\rm b2}})&\equiv &
2\int_{\theta}^{\theta_{{\rm b1}}}\!d\ln \theta~
\langle\gamma_+\rangle(\theta) \nonumber\\
&&+\frac{2}{1-\theta_{{\rm b1}}^2
/\theta_{{\rm b2}}^2}\int^{\theta_{{\rm b2}}}_{\theta_{{\rm b1}}}\!
d\ln\theta~ \langle\gamma_+\rangle(\theta)\nonumber\\
&=&\bar{\kappa}(<\theta)-\bar{\kappa}(\theta_{
 {\rm b1}}\le\theta\le\theta_{{\rm b2}}).
\label{eq:zeta}
\end{eqnarray}
Here, $\theta$ is a given radius and, $\theta_{{\rm b1}}$ and
$\theta_{{\rm b2}}$ are the inner and outer radii of subtracted
background region. The critical surface density $\Sigma_{\rm
cr}=c^2/(4\pi G) (D_{\rm s} {D_{\rm ls}/D_{\rm l}})$ is given by 
the angular diameter distances to cluster ($D_{\rm l}$), to source ($D_{\rm
s}$) and between cluster and source ($D_{\rm ls}$).
The $M_{\zeta_c}$ is a model-independent mass estimation.

We first obtain central positions of mass clumps by peak-finding
algorithms and then redistribute them within $0\farcm
2$ over 500 Monte Carlo simulations to take into account uncertainties
of central positions.  
We choice the central position at which the signal-to-nose ratio of
$\zeta_c$ measurement is at the maximum (Table \ref{tab:massclump}). 
The background region is in the annulus of $40-100$kpc surrounding candidates
so that we extract the substructure mass embedded in the cluster main potential.
If the cluster potential is uniform within $\sim100$kpc, it is a good mass
estimate of subclumps.
The following results are not changed by choosing the background region.
As described in eq. (\ref{eq:zeta}), $\zeta_c$ mass measurement is computed 
by integrating the measured tangential shears outside a given radius
$\theta$. 
Since the available number for
background galaxies in the outer annulus is larger than those in the
inner one, it enables us to plot $M_{\zeta_c}$ profile for each mass clump.

\begin{figure*}
\epsscale{1.0}
\plotone{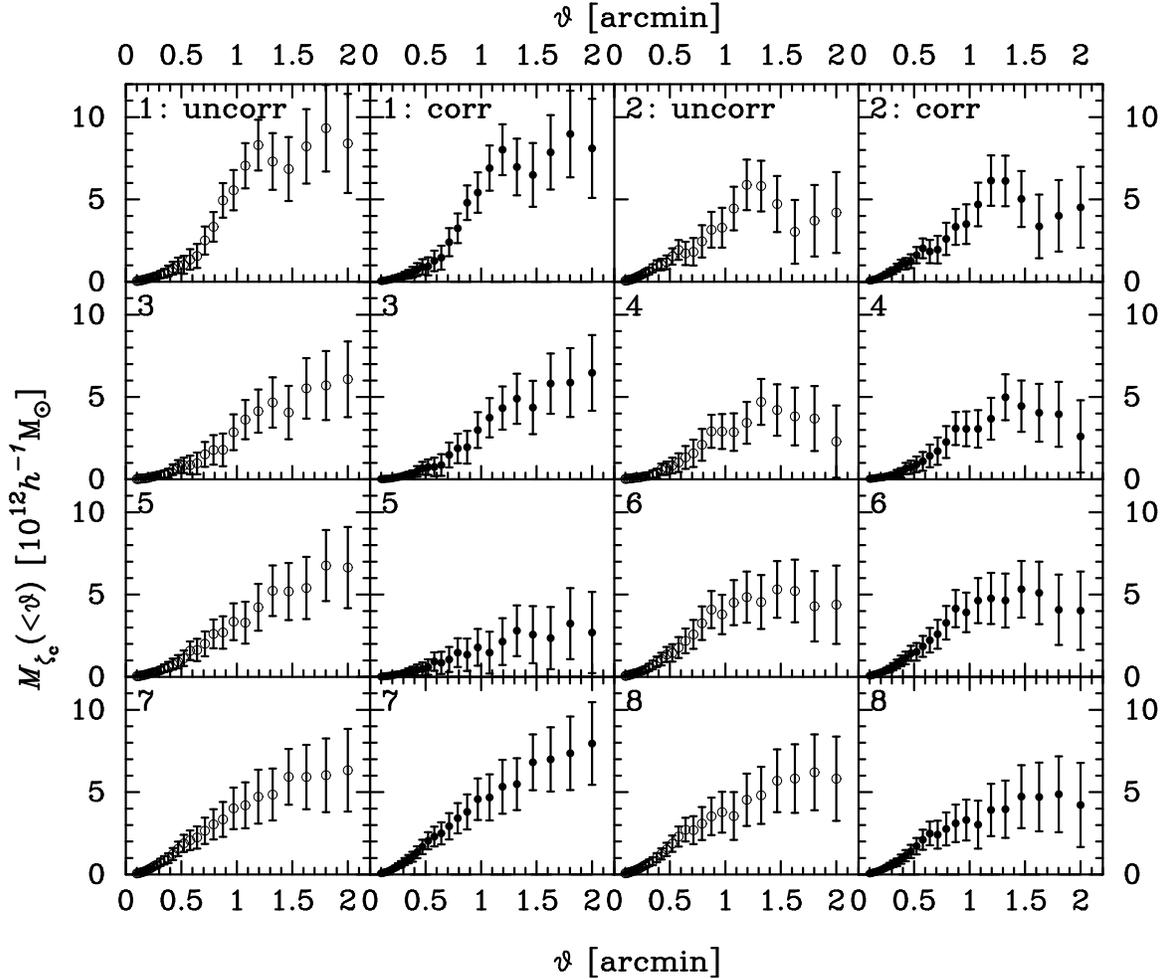}
\caption{ Aperture mass profile, $M_{\zeta_c}$, for each subclump
 candidate (1-8). Open and filled circles represent aperture masses without and
 with LSS correction, respectively.
Aperture masses are saturated on the outer radii, 
which indicates that the mass density in the surroundings are quite low. 
It is consistent with results of stacked lensing analysis
 (\S \ref{sec:stack}).
The aperture mass for candidate 5, corrected by LSS lensing effect,
 reduces to about half of that without a LSS correction.
}
\label{fig:Mzeta}
\end{figure*}

We also calculate LSS-corrected projected masses, $M_{\zeta_{\rm c}}^{\rm (corr)}$, 
in terms of $\langle
\gamma_+ \rangle(\theta)-\langle \gamma_+^{\rm (LSS)} \rangle(\theta)$,
where $\gamma_+^{\rm (LSS)}$ is the LSS shears obtained in \S \ref{subsec:proj}. 
We do not consider the intrinsic shape noise.
We note that $M_{\zeta_{\rm c}}^{\rm (corr)}$ is a model-dependent mass
because we assumed the mass-to-light ratio in a calculation of LSS
shears, $\gamma_\alpha^{\rm (LSS)}$. 
Figure \ref{fig:Mzeta} shows the LSS-uncorrected 
and corrected projected mass profiles, respectively.
The values of $M_{\zeta_c}$ are saturated on the outer radius, which 
indicates that the mass density of clumps is quite low on the outer radius. 
We estimate two-dimensional masses for each clump, $M_{\rm 2D}$ and $M_{\rm 2D}^{\rm (corr)}$, 
from the saturated values with a covariance matrix, 
because each bin is correlated with each other. 
We list the resulting masses in Table \ref{tab:massclump}.
The LSS- uncorrected and corrected masses are in good agreement with each other, but for candidate 5. 
The LSS-corrected mass for candidate 5 is about half of that before a correction. 
Therefore, the background LSS effect significantly contributes on
lensing signals for candidate 5.

The mean values are $\langle M_{{\rm
2D}}^{\rm (corr)}\rangle=(4.84\pm1.41)\times10^{12}h^{-1}M_\sun$ for all candidates
and $\langle M_{{\rm 2D}}^{\rm
(corr)}\rangle=(5.06\pm1.30)\times10^{12}h^{-1}M_\sun$ but for candidate 5.
They are at the order of the mass of cD galaxy halos, 
$M_{{\rm 3D}}^{\rm (SIS)}(\theta)=3.0\times10^{12}h^{-1}M_\sun (\sigma_v/400{\rm
km/s})^2 (r/40{\rm kpc})$, where $\sigma_v$ is the velocity
dispersion and we employ the velocity dispersion of cD galaxy
$\sim200-400{\rm km/s}$ (e.g. Smith et al. 2000).
Although we looked into the SDSS photometric and spectroscopic data, 
there is no correlation among the projected mass, and 
velocity dispersion and luminosity of galaxies which are located in each
subclump.

\begin{table*}
\caption{Projected Mass for Subclump candidates} \label{tab:massclump}
\begin{center}
\begin{tabular}{cccccccccc}
\hline
\hline
ID  & S/N
    & $\nu$
    & $P_{\rm spur}$
    & $b_{\rm map}$
    & $M_{\rm 2D}$ 
    & $M_{\rm 2D}^{\rm (corr)}$ 
    & (RA, Dec)
    & Name 
    & $d_{\rm off}$\\
    &
    &
    & $\%$
    &
    & $10^{12}h^{-1}M_\sun$
    & $10^{12}h^{-1}M_\sun$
    & deg
    & 
    & arcmin \\
(1) & (2)  & (3) & (4) & (5) & (6) & (7) & (8) & (9) & (10)\\
\hline
1  & $3.1$
   & $3.6$
   & $0.0$
   & $0.0$  
   & $8.01\pm1.05$ 
   & $7.70\pm1.05$
   & $(195.042, 27.996)$  
   & NGC 4889 
   & $1.2$
   \\
2  & $3.9$
   & $3.7$
   & -
   & -
   & $4.28\pm0.83$ 
   & $4.56\pm0.83$   
   & $(194.882, 27.948)$  
   & NGC 4874 
   & $1.2$
\\
3  & $3.2$
   & $2.9$
   & $0.4$
   & $1.3$
   & $5.03\pm0.87$ 
   & $5.29\pm0.87$
   & $(194.980, 27.843)$
   & SA 1656-054
   & $1.2$
\\
4  & $3.6$
   & $3.2$
   & $0.0$
   & $0.0$
   & $3.49\pm0.74$ 
   & $3.73\pm0.74$
   & $(194.722, 27.829)$
   & SDSS J125848.72+274837.5 
   & $0.7$
\\
5  & $3.4$
   & $3.1$
   & $12.2$
   & $41.1$
   & $5.95\pm1.08$ 
   & $2.74\pm1.08$
   & $(194.732, 27.722)$
   & -
   & -
\\
6  & $3.3$
   & $3.1$
   & $0.3$
   & $1.0$
   & $4.63\pm0.84$ 
   & $4.54\pm0.84$ 
   & $(194.724, 27.612)$
   & SDSS J125858.10+273540.9 
   & $1.4$
\\
7  & $3.1$
   & $3.1$
   & $0.1$
   & $0.3$
   & $6.07\pm1.24$
   & $7.32\pm1.24$
   & $(194.643, 27.625)$
   & NGC 4853 
   & $1.7$
\\
8  & $3.3$
   & $3.4$
   & $1.8$
   & $8.2$
   & $5.90\pm1.32$ 
   & $4.61\pm1.32$
   & $(194.475, 27.044)$
   & SDSS J125756.65+270215.0 
   & $0.8$
\\
\hline
\end{tabular}
\tablecomments{
Col. (1): Name of subclump candidate.
Col. (2): Signal-to-noise ratio in mass maps.
Col. (3): $\nu=\kappa/\kappa_{\rm rms}$. The rms $\kappa_{\rm rms}$ is
 obtained by 1000 Monte-Carlo simulations.
Col. (4) Probability of spurious lensing peak considering
contributions of main cluster, LSS effect and intrinsic noises.
Col. (5): Bias by LSS effect and intrinsic noise for a preference to detect a clump in
 the convergence field of main cluster, defined as $b_{\rm map}=P_{\rm spur}^{\rm (LSS)}/\langle P_{\rm spur}^{\rm
(noise)}\rangle$.
Col. (6): Projected mass for subclump without considering background LSS
 lensing effects. 
Col. (7): Projected mass for subclump with a correction of LSS lensing effects. 
Col. (8): Central positions for mass measurements.
Col. (9): Name of galaxies in the mass clump candidate.
Col. (10): Angular distance, $d_{\rm off}$, between luminous galaxy in
 clump candidate and centroid position for mass measurements.
We note that they are consistent with galaxy positions within 
the uncertainty of centroid position for clump candidates (${\rm FWHM}\simeq2\farcm$).
}
\end{center}
\end{table*}

\section{Tangential Shear Profile Stacked Over Five Subclumps} \label{sec:stack}

Since weak lensing signals of Coma cluster at the low
redshift is weak and the number of background galaxies within small
radius is few, it is quite difficult to measure tangential shear profile
for each subclump candidate. We therefore measure a profile of
tangential shear components ensembling five subclump candidates.
A mass measurement with stacked tangential distortion profile is
complementary to the $\zeta_c$ statistics (\S \ref{sec:submass}), because different shear catalogue is
used in two measurements.
In the $\zeta_c$ statistics, source galaxies outside a given
radius is used, while, in the tangential shear measurement, source galaxies from
inner to outer radius is independently available.
We here exclude two dark halos associated with cD galaxies 
in order to
avoid a contamination of lensing distortion caused by the main cluster.
The candidate 5 at which the projection effect is significant is
also ignored.
The center for each subclump is chosen as the same position of the $M_{\zeta_c}$
measurements (\S \ref{sec:submass}).  
The averaged tangential shear distortions of source galaxies, $\langle g_+ \rangle
(\theta_n)$, is calculated in the circular annulus of the same radius, 
based on the same procedure as \S\ref{sec:mass}. 
The typical projected distance between a center of a stacked tangential shear
profile and the main cluster center, $\langle \theta_{{\rm off}} \rangle$, is
obtained with a weight function of lensing signals,
$\langle \theta_{{\rm off}} \rangle^2=\sum_j \langle g_{+,j} \rangle^2 \theta_{{\rm off},j}^2/ \sum_j
\langle g_{+,j} \rangle ^2\sim17\farcm4$,
where $\langle g_{+,j} \rangle$
is the lensing signal (\S \ref{sec:mass}) for each subclump and $\theta_{{\rm off},j}$ is
an angular separation between each subclump and the main center.

\begin{figure*}
\plottwo{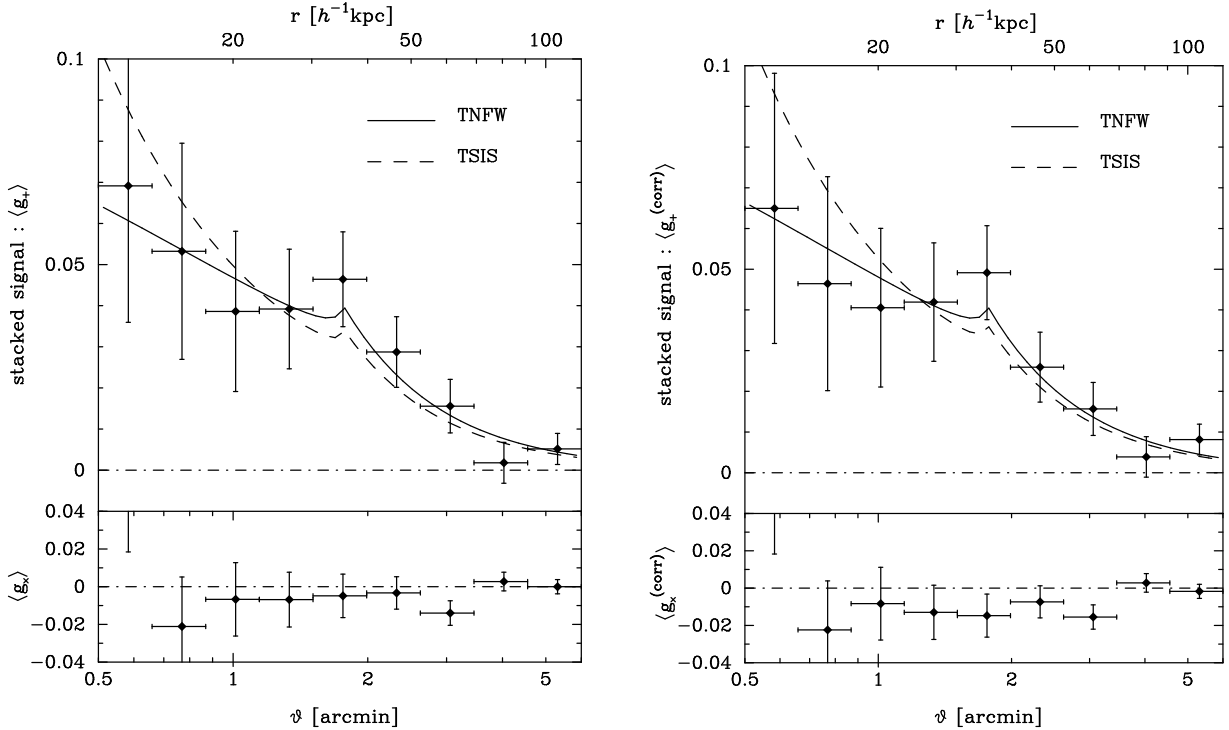}{f9b.ps}
\caption{Profiles of tangential shear component (top panel), $g_+$, and the $45$ degree rotated
component (bottom panel), $g_\times$, obtained from an ensemble of five subclumps.
{\it Left} : the stacked profile from original shear catalogue. {\it Right} :
 the profile with a LSS lensing correction.
The $g_+$ values clearly decrease over $\theta\sim1\farcm75$.
The solid and dashed lines are subhalo models of TSIS and TNFW, respectively.
}
\label{fig:stack}
\end{figure*}

We compute the LSS-corrected shear profile, $\langle g_+
\rangle(\theta_n)-\langle g_+^{\rm (LSS)} \rangle(\theta_n)$, where the azimuthal average of
the LSS distortion components, $\langle
g_+^{\rm (LSS)} \rangle(\theta_n)$, are calculated without statistical weight ($u_{g,i}=1$).
Figure \ref{fig:stack} shows the stacked tangential shear profiles as a
function of transverse separation $\theta$, 
with and without the LSS effect. 
We estimate the contribution of the main cluster mass on the stacked
lensing signals, because the tangential shear provides full information on the lensing
signals of gravitational potentials of both the main cluster and subclumps.
It is necessary to calculate lensing distortions caused by the main
cluster in order to measure typical mass of interior subhalos.
We follow the convolution technique of Yang et al. (2006) to measure 
the azimuthally averaged convergence at the offset from the main cluster
center in the lens plane. 
The values of $g_+$ of the offset main halo is at the order of ${\it
O}(-10^{-5})$ in the range of the stacked lensing profile ($\theta<6\farcm$), 
which indicates that the lensing signals from the main cluster at the positions of 
subclumps are negligible.

Since the tidal field of main cluster disrupts dark matter halos of
interior substructures, the subhalo radius would be determined by the
tidal radius rather than the virial radius $r_{\rm vir}$.
We therefore consider a truncated SIS model (TSIS) and a truncated NFW
model (TNFW) for the tangential fitting, whose density profiles are
truncated at the radius $r_t$.
\begin{eqnarray}
  \rho_{\rm TSIS}(r) &=& \frac{\sigma_{v,t}^2}{2\pi G r^2}~~~~{\rm
   for}~~r \le r_t \label{eq:rhoTSIS}  \\
                  &=& 0      ~~~~~~~~~~~{\rm for}~~r > r_t  \nonumber \\
  \rho_{\rm TNFW}(r) &=& \frac{M_{\rm sub}^{\rm (TNFW)}}{4\pi r_{s,t}^3 m(c_t)}\frac{1}{(r/r_{s,t})(1+r/r_{s,t})^2}~~~{\rm for}~~r \le r_t  \nonumber  \\
                  &=& 0~~~~~~~~~~~~~~~~~~~~~~~~~~~~~~~~~~~{\rm for}~~r > r_t, \label{eq:rhoTNFW}
\end{eqnarray}
where $\sigma_{v,t}^2$ is a velocity dispersion for TSIS
model, $M_{{\rm sub}}^{\rm (TNFW)}$ and $c_t$ are a mass and a concentration for
TNFW model, and $r_{s,t}=r_t/c_t$ is a scale radius determined by the
concentration and the truncated radius $r_t$. The subclump mass for TSIS
model is given by $M_{\rm sub}^{\rm (TSIS)}=2\sigma_{v,t}^2r_t/G$.

Analytical expressions of the two-dimensional projection $\Sigma$ of the
density field are obtained by an integration over $r_\parallel=[-\sqrt{r_t^2-r^2}, +\sqrt{r_t^2-r^2}]$
\begin{eqnarray}
\kappa_{\rm
 TSIS}(\theta) &=&
\left\{ \begin{array}{ll}
\frac{1}{\pi}\left(\frac{\theta_{E,t}}{\theta}\right)\arccos\left(\frac{\theta}{\theta_t}\right)
& ~~{\rm for}~~ \theta \le \theta_t \nonumber \\
0 
& ~~ {\rm for}~~ \theta>\theta_t
\end{array} \right.
\end{eqnarray}
where $\theta=r/D_l$ is an angular size of the three dimensional radius, 
and $\theta_{E,t} \equiv 4\pi (\sigma_{v,t}/c)^2D_{\rm ls}/D_{\rm s}$ is
the Einstein radius for TSIS model. 
\begin{eqnarray}
\kappa_{\rm  TNFW}(\theta) &=& \frac{M_{\rm sub}^{\rm (TNFW)}}{2\pi
\Sigma_{\rm cr} r_{s,t}^2 m(c_t)} f(x),~~~~x=\theta/\theta_{s,t}\\
f(x)&=&\left\{ \begin{array}{l}
-\frac{\sqrt{c_t^2-x^2}}{(1-x^2)(1+c_t)}+\frac{1}{(1-x^2)^{3/2}}
{\rm arccosh} \left(\frac{x^2+c_t}{x(1+c_t)}\right) \\ 
{\rm~~~~~~~~~~~~~~~~~~~~~~~~~~~~~~for}~~x < 1 \\
\frac{\sqrt{c_t^2-1}}{3(1+c_t)} \left(1+\frac{1}{1+c_t}\right)  \\
{\rm~~~~~~~~~~~~~~~~~~~~~~~~~~~~~~for}~~x = 1  \\
-\frac{\sqrt{c_t^2-x^2}}{(1-x^2)(1+c_t)}-\frac{1}{(x^2-1)^{3/2}}{\rm arccos}\left(\frac{x^2+c_t}{x(1+c_t)}\right)
 \\ {\rm~~~~~~~~~~~~~~~~~~~~~~~~~~~~~~for}~~ 1 < x \le c_t \\
0  \\ {\rm~~~~~~~~~~~~~~~~~~~~~~~~~~~~~~for}~~ c_t < x 
\end{array} \right.
\end{eqnarray}
The expression of $f(x)$ is the same as Takada \& Jain (2003) and Hamana, Takada
\& Yoshida (2004), although they adopt that the mass density is
truncated at the virial radius.
We here do not assume $r_t=r_{\rm vir}$, because we aim to investigate disrupted
interior substructures. The TSIS and TNFW profiles are specified in
terms of two parameters of $\theta_{E,t}$ and $\theta_t$ and three
parameters of $M_{\rm sub}^{\rm (TNFW)}$, $c_t$ and $\theta_t$, respectively.

We fit the LSS- uncorrected and corrected distortion profiles with TSIS and TNFW models.
The best-fit parameters are summarized in Table \ref{tab:fit}.
Two models well describe the stacked tangential shear profile.
The best-fit parameters with and without a LSS lensing correction does
not change significantly.
The truncated radii for two models are in good agreements with each other.
Indeed, the break in the tangential shear profile is clearly found at
$\theta\sim1\farcm75$ (Figure \ref{fig:stack}). The values of $g_+$
steeply decrease ($\propto \theta^{-2}$) over the truncated radius,
which indicates that the halo mass density drops to zero at the
truncated radius. It is consistent with the $\zeta_{\rm c}$ mass measurement (\S
\ref{sec:submass}). 
This feature does not clearly appear in a stacked lens
analysis for massive clusters of $M_{\rm vir}>10^{14}h^{-1}M_\sun$ (Okabe et al. 2009).
The TNFW and TSIS masses are in agreements with the
the mean projected mass $\langle M_{\rm 2D}^{\rm (corr)}
\rangle_{34678}=(4.78\pm1.07)\times10^{12}h^{-1}M_\sun$. 
We note that the two-dimensional mass for truncated mass models (TNFW
and TSIS) has the same analytical expression as the three-dimensional
mass ($M_{\rm 2D}^{{\rm TSIS}}=M_{\rm 3D}^{{\rm TSIS}}$ and $M_{\rm
2D}^{{\rm TNFW}}=M_{\rm 3D}^{{\rm TNFW}}$), because there is no
projection effect due to the zero mass outside $r_t$.

We also investigate model fittings for spurious mass clumps appeared in mass maps
simulated by rotated shear catalogue. 
First, 
we randomly rotate an angle in the ($g_1, g_2$) plane for each background galaxies
with it's $|g|$ fixed and then conduct the mass reconstruction for
background catalogue by 500 times.
Second, we detect peaks whose significance is over $3\sigma$ in mass
maps. Third, we measure stacked tangential profiles for 5 peaks which are
bootstrap re-sampled from detected peaks by 300 times.  
In the measurements, the peak whose distance
from cluster center is more than $10\farcm$ is only used in order to
avoid the main cluster lensing signal (Yang et al. 2006). 
The averaged $\chi^2$s are $\langle \chi^2\rangle=9.9$ for TNFW and 
 $\langle \chi^2\rangle=15.7$ for TSIS models.
Those $\chi^2$s are worse than our results.
This is why the form of stacked
tangential profile for spurious clumps is different from Figure
\ref{fig:stack}.

\begin{table}
\caption{The best-fit truncated SIS (TSIS) and truncated NFW (TNFW)
 model parameters.} \label{tab:fit}
\begin{center}
\begin{tabular}{ccc}
\hline
\hline
TSIS & LSS uncorrected & LSS corrected\\
(1)  & (2) & (3) \\
\hline
$M_{\rm sub}^{\rm (TSIS)}$ [$10^{12}h^{-1}M_\sun$] & $3.47_{-0.56}^{+1.21}$ & $3.54_{-0.54}^{+1.32}$\\
$\theta_t$ [arcmin] & $1.76_{-0.15}^{+0.57}$ & $1.76_{-0.13}^{+0.61}$          \\
$r_t$ [$h^{-1}{\rm kpc}$] & $35.21_{-2.98}^{+11.32}$ & $35.21_{-2.58}^{+12.25}$ \\ 
$\sigma_{v,t}$ [{\rm km/s}] & $460.59_{-32.70}^{+30.53}$ &  $465.27_{-32.31}^{+30.21}$\\ 
$\chi^2/{\rm d.o.f}$ &  $4.4/7$  & $5.3/7$ \\
\hline
TNFW & LSS uncorrected & LSS corrected \\
(4)  & (5) & (6) \\
\hline
$M_{\rm sub}^{\rm (TNFW)}$ [$10^{12}h^{-1}M_\sun$] & $3.90_{-0.47}^{+1.22}$ &  $4.00_{-0.37}^{+1.45}$ \\
$\theta_t$ [arcmin] & $1.75_{-0.26}^{+0.25}$ & $1.75_{-0.29}^{+0.21}$ \\
$r_t$ [$h^{-1}{\rm kpc}$] & $35.01_{-5.28}^{+4.99}$ & $35.01_{-5.79}^{+4.21}$ \\
$c_t$ & $1.53_{-0.79}^{+1.12}$ & $1.53_{-0.68}^{+0.74}$ \\
$\chi^2/{\rm d.o.f}$ &  $2.5/6$  & $2.4/6$\\
\hline
\end{tabular}
\tablecomments{
Col. (1): Name of best-fits parameters for TSIS model
 (eq. \ref{eq:rhoTSIS}) 
: the subhalo mass $M_{\rm sub}^{\rm (TSIS)}$, 
the truncated radius, $r_t$, its angular radius $\theta_t$ and the
 velocity dispersion $\sigma_{v,t}$. 
Col. (2-3) : The best-fit parameters for TSIS model without and with LSS correction.
Col. (4) : Name of best-fits parameters for TNFW model (eq. \ref{eq:rhoTNFW}):
the subhalo mass $M_{\rm sub}^{\rm (TNFW)}$, 
the truncated radius, $r_t$, its angular radius $\theta_t$ and the
 concentration parameter $c_{t}$. 
Col. (5-6) : The best-fit values for TNFW model without and with LSS correction.
All parameters are determined by fitting the mean distortion profile
 which is obtained by staking the distortion signals for 5 clump candidates
 (3,4,6,7 and 8).
The $\chi^2/{\rm d.o.f}$ is the chi-square for best-fits and the degree-of-freedom.
}
\end{center}
\end{table}

\section{Discussion and Conclusion}  \label{sec:dis}

\subsection{Cluster Mass Comparison}
We compare our mass estimates with previous results of
multi-wavelength data. 
The line-of-sight velocity distribution of member galaxies with the
Jeans equation which requires the assumption of the dynamical
equilibrium derived $M_{\rm vir}=9.8\times 10^{14}h^{-1} M_\sun$
and $c_{\rm vir}=9.4$ for NFW mass model (\L okas \& Mamon 2003).
The caustic mass estimates to use a characteristic pattern 
in the redshift-space, formed by galaxies falling into cluster potential 
(Kaiser 1987) obtained $M_{\rm 200}=7.85\times 10^{14}h^{-1} M_\sun$ and
$c_{200}=10$ (Rines et al. 2003). 
Our result of NFW mass $M_{\rm vir}$ and $M_{\rm 200}$ is in good agreement with
their results, while our concentration parameter is lower.
Our weak lensing analysis is to use one band imaging
data. As demonstrated by Broadhurst et al. (2005), see also ( Umetsu \&
Broadhurst 2008 ; Okabe et al. 2009), the dilution contamination of member galaxies on lensing
signals is problematic for an accurate measurement of the concentration
parameter. It is therefore of importance to correct the dilution effect by 
secure selecting of background galaxies in the color-magnitude plane
(Broadhurst et al. 2005; Umetsu \& Broadhurst 2008; Umetsu et al. 2009).
In addition, we require the data, which covers the virial radius, in
order to improve measurement accuracy of halo mass.
The SDSS and CHFT weak lensing results (Kubo et al. 2007; Gavazzi et
al. 2009) are
$M_{200}=18.8^{+6.5}_{-5.6}\times 10^{14}M_\sun$ and
$M_{200}=5.1^{+4.3}_{-2.1}\times 10^{14}M_\sun$,
  and $c_{200}=3.84^{+13.16}_{-0.18}$ and $c_{200}=5.0^{+3.2}_{-2.5}$.
They do not disagree with our results within large errors. 
The ASCA and ROSAT X-ray observations with assumptions of hydrostatic equilibrium,
isothermally and single $\beta$ model shows 
$M_{500}=11.99^{+1.28}_{-1.29}\times 10^{14}M_\sun$ (Reiprich \& B\"ohringer 2002)
and 
$M_{500}=9.95^{+2.10}_{-2.99}\times 10^{14}M_\sun$ (Chen et al. 2007),
which are higher than our estimates.
We cannot rule out a possibility that the low angular resolution of
ASCA satellite leads to a bias on mass estimates.
It is of critical importance to compare X-ray and weak-lensing
masses of Coma cluster which is the only cluster known to have
turbulence in the ICM.
The ASCA and XMM-Newton X-ray observations (e.g. Watanabe et al. 1999;
Arnaud et al. 2001) have shown the complex temperature variations in the
intra-cluster medium (ICM). Schuecker et al. (2004) has revealed 
a Kolmogorov/Oboukhov-type turbulence spectrum in the ICM as a
consequence of the projected pressure distributions.
They constrained that the lower limit of turbulent pressure accounts for
10 percent of the total ICM pressure. 
Recent hydrodynamic N-body simulations pointed out that X-ray
mass estimates with an assumption of hydrostatic equilibrium 
are biased low due to ICM turbulence,
because the gas motion pressure of turbulence as well as bulk motions supports
a part of the total pressure (e.g. Ervard et al. 1999; Nagai et al. 2007). 
Their results cast a doubt on accurate cluster mass measurement by X-ray
analysis alone, which is a serious concern for cluster-based
cosmological probes (e.g. Vikhlinin et al. 2009a,b; Okabe et al. 2010; Zhang et al. 2010). 
Therefore, a comparison of independent mass estimates is of great
importance to understand, in a quantitative manner, how much the gas motion
pressure affects the X-ray mass estimates.

\subsection{Comparison with X-ray image}

\begin{figure}
\plotone{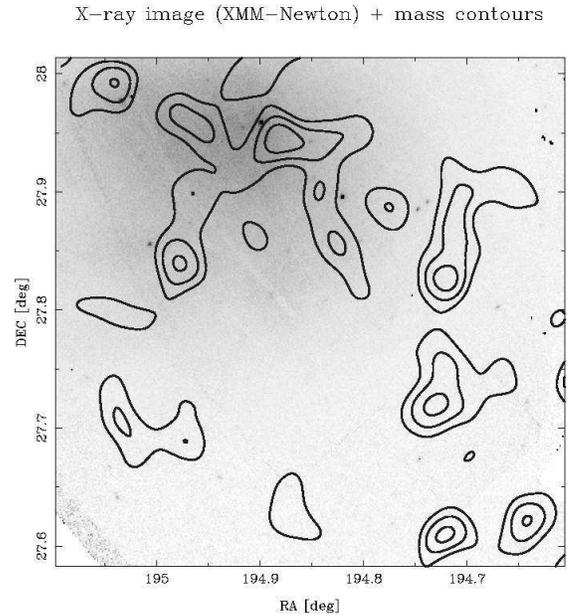}
\caption{X-ray exposure-corrected image, no-subtracted backgrounds,  overlaid with mass contours of the central $26\farcm
 \times 26\farcm$ cluster region.}.
\label{fig:xmm}
\end{figure}

We compare an X-ray exposure-corrected image retrieved from
archival data of XMM-Newton with mass counters of the central region
(Figure \ref{fig:xmm}). The XMM-Newton data in the outskirts region is
lacked. The X-ray image shows some
point sources associated with galaxies in clump candidates. The intra-cluster-medium (ICM)
distributions are not correlated with mass clump candidates. This is why
the intra-cluster plasma can be escaped from the gravitational potential of
subclumps. The sound velocity of the ICM is given by
\begin{eqnarray}
c_s=\left( \frac{5k_BT}{3\mu m_p}\right)^{1/2}\simeq1457\left(\frac{k_BT}{8.25{\rm
keV}}\right)^{1/2}{\rm km/s},
\end{eqnarray}
where the temperature $k_BT=8.25\pm0.10$keV (Arnaud, et al. 2001), 
the mean molecular weight $\mu=0.62$, and the proton mass $m_p$.
The sound velocity is higher than the typical escape velocity from subclumps, as below
\begin{eqnarray}
c_s \ge v_{\rm esc}=\left(\frac{2 G M_{\rm sub}^{\rm (TNFW)}}{
	      r_t}\right)^{1/2}\simeq991{\rm km/s}.
\end{eqnarray}
Since the temperature of point sources is low $k_BT\sim1{\rm keV}$ and
high metal abundance (Vikhlinin et al. 2001), it's sound velocity is lower than the escape
velocity.

\subsection{Tidal Radius}

The stacked tangential shear profile for five subclumps, excluding subhalos
associated with cD galaxies and candidate 5, is well described by the
TSIS and TNFW  models.
The fitting result gives the truncated radius
$r_t\simeq 35h^{-1}~{\rm kpc}$, which coincides with 
galaxy-galaxy lensing results in clusters (e.g. Natarajan \& Springel 2004;
Natarajan, De Lucia  \&  Springel 2007).
The truncated radius is much smaller than a truncated radius $\sim
200~{\rm kpc}$ obtained by galaxy-galaxy lensing studies of fields
(e.g. Hoekstra, Yee \&  Gladders 2004). 
It would be due to the strong tidal field of the main cluster
gravitational potential.
The tidal radius of a subhalo orbiting in a spherically,
symmetric mass distribution of a cluster is obtained  by the balance between
the tidal force of the primary halo and the gravity of subclump,
$r_{\rm tidal}=(M_{\rm sub}/(M(<r_p)(2-\partial \log M/ \partial \log r_p
))^{1/3}r_p$
(e.g. Tormen et al. 1998),
where $r_p$ is the pericenter radius which is
the minimal radius from the cluster center during its orbiting history.
Since we do not constrain its pericenter radius directly from a current
position and not derive the three-dimensional radius, 
we instead assume the mean, projected offset radius $\langle \theta_{\rm
off} \rangle $ of subclumps in stacked lensing analysis.
Here, we assume the NFW and TNFW mass models for the main cluster
and substructures, respectively.
We obtain the tidal radius $r_{\rm
tidal}\sim42~h^{-1}{\rm
kpc}$, which coincides with the truncated radius $r_t\simeq 35
~h^{-1}{\rm kpc}$.

\subsection{Subhalo Mass Fraction}

We found four subhalos and two cD galaxy halos in the central data
($r\simlt 30\farcm$) and one halo in the outskirts data ($30\farcm
\simlt r\simlt 60\farcm$).
There is a difference of the halo number for the radius, which might
support the results of numerical simulation (e.g De Lucia et al. 2004;Gao et
al 2004) that the number of substructure increases as the radius decreases.
Compensating the limitation of our data region, we roughly estimate the number
of subhalos within the radii $r_{\rm vir}$ and $r_{200}$.
If we assume that there are four ($r\simlt 30\farcm$) and one ($30\farcm
\simlt r\simlt 60\farcm$) subhalos in the area corresponding to our data, the halo number 
is estimated to be $\pi (r_{\rm vir}^2-30\farcm ^2)/A_{\rm
outskirts}+4\times (\pi 30\farcm^2/A_{\rm center})+2 {\rm cDs}$, 
where $A$ is the area of our data. 
The Poisson noise for distributions is applied for the statistical errors.

The halo number detected by weak lensing analysis 
is expected to be $N_{\rm sub}(<r_{\rm vir})=43\pm30$ and $N_{\rm
sub}(<r_{\rm 200})=27\pm15$. 
 If the typical halo mass is the best-fit value $\langle M_{\rm
sub}^{\rm (TNFW)}\rangle=4.00\times10^{12}h^{-1}M_\sun$ obtained
from fitting of the stacked tangential profile, 
the total substructure mass within the virial radius account for
 $\sim19\pm13$ and $17\pm9$ 
percents of total cluster masses $M_{\rm vir}$ and $M_{200}$, respectively. 
Although the total mass fraction contained in subhalos does not agree each other among 
literature (e.g. De Lucia et al. 2004; Natarajan, De Lucia
\&  Springel 2007, Gao et al. 2004), most authors estimate 5 -20 percent.
Our result is in rough agreement with the numerical simulations.
A galaxy-galaxy lensing study (Natarajan, De Lucia \&  Springel 2007) indicates that
$10-20\%$ of the mass is contained in cluster substructures, which
also roughly agrees with our result.

Our weak lensing analysis on the nearby cluster would indicate the
possibility that the mass function of cluster substructures is
measurable without assumptions of mass-to-light ratio for member
galaxies and dynamical state, while the galaxy-galaxy lensing
studies (e.g. Natarajan \& Springel 2004; Natarajan, De Lucia  \&
Springel 2007) requires the assumption of the mass and light scaling law.
We however have not yet obtained the mass spectrum as the galaxy-galaxy
lensing studies (e.g. Natarajan \& Springel 2004; Natarajan, De Lucia
\&  Springel 2007).

Alternative possible approach to investigate the mass function of cluster
substructures is to measure higher order moments of the lensed images
(HOLICs) and using the moments to estimate the flexion (e.g. Okura et
al. 2007; Okura \& Futamase 2008; Okura et al. 2008).
They have shown for the first time that flexion analysis can discover
substructures using the image of A1689 (Okura et al. 2007).

As pointed out by Shaw et al. (2006), the median mass fraction is a
increase function of the virial mass ($f_{\rm sub}\propto
M_{\rm vir}^{0.44\pm0.06}$), because massive objects, which formed more recently
than less massive objects, have less time to disrupt subhalos (Zenter et
al. 2005). 
The statistical study of mass fraction of galaxy clusters
is, therefore, one of good tests of $\Lambda$CDM and the hierarchical
clustering, as the concentration-mass relation of the NFW mass model
(Okabe et al. 2009).
Hence, further systematic study of mass fractions is required.

The area of our current data is insufficient to derive the halo mass
function as well as to measure cluster mass accurately.
The next instrument of a prime focus camera of Subaru telescope,
Hyper-Suprime-Cam, whose field-of-view is $\sim1.5 {\rm deg^2}$, 
will efficiently observe the nearby cluster and enables us to conduct
weak lensing and flexion analyses. 
Our result using the Subaru/Suprime-Cam does guarantee that weak lensing analysis using 
Subaru/Suprime-Cam and Hyper-Suprime-Cam is capable for almost X-ray
clusters.

\section*{Acknowledgments}

We gratefully thank the anonymous referee whose comments significantly improved the manuscript. 
We are grateful to N. Kaiser for developing the IMCAT package
 publicly available. 
We thank Gavazzi, R. for discussing a CHFT weak lensing analysis.
N.O. gratefully thanks H. Hayashi, Y. Itoh, M. Chiba, M. Takada,
K. Umetsu and H. Nishioka for helpful discussions.
N.O. and T.F. are in part supported by a Grant-in-Aid from the
Ministry of Education, Culture, Sports, Science, and Technology of Japan
(NO: 20740099; TF: 20540245) as well as the GCOE program 
 ``Weaving Science Web beyond Particle-matter Hierarchy'' at Tohoku
 University and 
a Grant-in-Aid for Science Research in a Priority Area "Probing the Dark
Energy through an Extremely Wide and Deep Survey with Subaru Telescope"
(No. 18072001).
Y.O. thanks the JSPS Research Fellowships for Young Scientists.

\clearpage

\end{document}